\documentclass{aa}  

\usepackage{graphicx}
\usepackage{txfonts}
\usepackage[colorlinks=true]{hyperref}
\usepackage{xcolor,colortbl}
\usepackage{float}
\usepackage{adjustbox}
\usepackage{pdflscape}
\usepackage{lscape}
\usepackage{comment}
\usepackage{changepage}
\usepackage{geometry}
\usepackage{threeparttable}
\hypersetup{linkcolor=blue,citecolor=blue}
\makeatletter
\renewcommand*\aa@pageof{, page \thepage{} of \pageref*{LastPage}}
\makeatother
\usepackage{diagbox}
\usepackage{multirow}
\usepackage{graphicx} 
\graphicspath{ {./images/} }

\usepackage[rightcaption]{sidecap}

\usepackage{wrapfig}
\usepackage{tabularx}
\usepackage[caption=false]{subfig}

\hypersetup{urlcolor=blue}

\begin{document}

   \title{A versatile classification tool for galactic activity using optical and infrared colors}

   \author{C. Daoutis
          \inst{1,2}
          \and
          E. Kyritsis\inst{1,2}
          \and 
          K. Kouroumpatzakis \inst{3,2,1}
          \and
          A. Zezas \inst{1,2,4}
          }

   \institute{ Physics Department, and Institute of Theoretical and Computational Physics, University of Crete, 71003 Heraklion, Greece \\
         \email{cdaoutis@physics.uoc.gr}
         \and
             Institute of Astrophysics, Foundation for Research and Technology-Hellas, 71110 Heraklion, Greece
        \and 
            Astronomical Institute, Academy of Sciences, Bo\v{c}ní II 1401, CZ-14131 Prague, Czech Republic
        \and
            Center for Astrophysics | Harvard \& Smithsonian, 60 Garden St., Cambridge, MA 02138, USA
            }

   \date{Received September 15, 1996; accepted March 16, 1997}

 
  \abstract
   {The overwhelming majority of diagnostic tools for galactic activity are focused mainly on the classes of active galaxies. Passive or dormant galaxies are often excluded from these diagnostics which usually employ emission line features (e.g., forbidden emission lines). Thus, most of them focus on specific types of activity or only on one activity class (e.g., active galactic nuclei or AGN).}
   {In this work, we use infrared and optical colors in order to build an all-inclusive galactic activity diagnostic tool that can discriminate between star-forming, AGN, LINER, composite, and passive galaxies, and which can be used in local and low-redshift galaxies.}
   {We use the random forest algorithm in order to define a new activity diagnostic tool. As ground truth for the training of the algorithm, we consider galaxies that have been classified based on their optical spectral lines. We explore classification criteria based on infrared colors from the first 3 \textit{WISE} bands (bands 1, 2, and 3) supplemented with optical colors from the \textit{u}, \textit{g}, and \textit{r} SDSS bands. From these we seek the combination with the minimum number of colors that provides optimal results. Furthermore, to mitigate biases related to aperture effects, we introduce a new \textit{WISE} photometric scheme combing different sized apertures.}
   {We develop a diagnostic tool using machine learning methods that accommodate both active and passive galaxies under one unified classification scheme using just 3 colors. We find that the combination of W1-W2, W2-W3, and \textit{g}-\textit{r} colors offers good performance while the broad availability of these colors for a large number of galaxies ensures wide applicability on large galaxy samples. The overall accuracy is $\sim$81\% while the achieved completeness for each class is $\sim$81\% for star-forming, $\sim$56\% for AGN, $\sim$68\% for LINER, $\sim$65\% for composite, and $\sim$85\% for passive galaxies.}
   {Our diagnostic provides a significant improvement over existing IR diagnostics by including all types of active, as well as passive galaxies, and extending them to the local Universe. The inclusion of the optical colors improves their performance in identifying low-luminosity AGN which are generally confused with star-forming galaxies, and helps to identify cases of starbursts with extreme mid-IR colors which mimic obscured AGN galaxies, a well-known problem for most IR diagnostics.}


   \maketitle
%

\section{Introduction}

Galaxies can be classified into different categories based on their activity. Some form new stars (i.e., star-forming galaxies, also referred to as \ion{H}{II} galaxies due to their \\ion{H}{II} region-like spectra), while other ones may present intense nuclear activity fueled by their supermassive black hole at their nucleus (AGN). In addition, galaxies can present, simultaneously, both of these behaviors. These are known as composite galaxies or transition objects \citep[TO, e.g.,][]{1997ApJ...487..568H}. In another galactic category, we find galaxies that host old stellar populations, contain small amounts of gas or dust and they do not exhibit any star formation or nuclear activity. These are the passive galaxies. Finally, there is also the LINER \citep[low-ionization nuclear emission-line region;][]{Heckman} galaxies. These galaxies can be separated into two distinct categories: LINERs which are powered by a supermassive black hole \citep[LINER type 1;][]{1997ApJ...487..568H} and
LINERs in which the source of excitation is the UV emission from post-AGB stars \citep{1994A&A...292...13B,2008MNRAS.391L..29S,2013A&A...555L...1P}.

Until now, the best way to discriminate between the four classes of active galaxies mentioned above (star-forming, AGN, LINER, and composite) is through the emission-line ratio diagrams introduced by \cite{1981PASP...93....5B}, hereafter BPT diagram. These are 2-dimensional diagrams that separate galaxies into \ion{H}{II} regions (star-forming), AGNs (Seyfert), LINERs, and composites using the characteristic emission-line ratio fluxes. The most commonly used version of this diagram is a plot of [\ion{O}{III}]$\lambda$5007/H$\beta$ against either [\ion{N}{II}]$\lambda$6584/H$\alpha$ or one of the [\ion{S}{II}]$\lambda \lambda$6716,6731/H$\alpha$ or [\ion{O}{I}]$\lambda$6300/H$\alpha$ \citep{2001ApJ...556..121K,2003MNRAS.346.1055K,2007MNRAS.382.1415S}. The classification of a galaxy depends
on its location on those diagrams. Although this has been a highly accurate and reliable method for galactic activity classification for many years, it presents some disadvantages. One of them is that in order to classify a galaxy, one needs to obtain an optical spectrum which can be challenging for very large samples of galaxies. A second reason is absorption by the interstellar medium (ISM) which may obscure the AGN emission. Additionally, some of the emission lines are weak hampering the application of these diagnostics to faint objects. In order to overcome these difficulties, new methods for classifying galaxies have emerged using infrared photometry and more specifically in the mid-infrared (3-24 $\mu$m) part of the spectrum. The use of photometry allows the application of the diagnostic to large samples of galaxies, while the use of IR data allows the identification of obscured AGN. 

Observations with the Spitzer Space Telescope \citep{2004ApJS..154....1W} led to the development of the first versatile activity diagnostics in the near- to mid-IR by \cite{2005ApJ...631..163S} and \cite{2012ApJ...748..142D}.
Subsequently, the launch of the \textit{WISE} satellite \citep[\textit{Wide-field Infrared Survey Explorer};][]{2010AJ....140.1868W}, enabled systematic studies of large populations of galaxies by providing sensitive all-sky photometry in the 3-24 $\mu$m (wavelength range in four bands, W1, W2, W3, and W4 having effective wavelengths of 3.4 $\mu$m, 4.6 $\mu$m, 12.0 $\mu$m, and 22.0 $\mu$m respectively). This led to the development of a new family of diagnostic tools.

One widely used diagnostic for AGN identification based on \textit{WISE} infrared photometry is the criterion of W1-W2 $\geq$ 0.8 \citep{2012ApJ...753...30S}. Another diagnostic is based on the W1-W2 color against the W2-W3 color \citep{2012MNRAS.426.3271M}.

Even though, these two infrared AGN selection methods have had great success in identifying high redshift galaxies in several surveys \citep[e.g., CANDELS;][]{2011AAS...21832803K}, they are tailored toward higher-redshift, more luminous, or obscured AGN. In fact, the diagnostic of \cite{2012MNRAS.426.3271M} is built based on an X-ray selected sample of AGN. However, application of such diagnostics to other samples of galaxies shows that they fail to identify a large population of AGN, especially in the local Universe. In a sample of galaxies taken from the Sloan Digital Sky Survey (SDSS), most of the AGN galaxies are located below the W1-W2 = 0.8 AGN selection line of \cite{2012ApJ...753...30S} or are located outside the AGN wedge of \cite{2012MNRAS.426.3271M} (see Sec. \ref{comp_mat}).

In order to overcome this limitation, we develop a new mid-IR-optical color activity diagnostic using advanced methods like machine learning algorithms aiming to supplement and enhance the performance of the existing diagnostic methods. The main reason we consider these algorithms as the basis of our diagnostic tool is that there is strong mixing between the mid-IR colors of the different types of galaxies. The availability of sensitive all-sky surveys provide photometric data for millions of galaxies. Machine learning methods allow us to efficiently exploit these rich databases and capture their complexity in multi-dimensional parameter spaces.

Since in this work, we do not use emission lines, we are able to include the class of passive galaxies, which is often excluded in standard diagnostic tools. Therefore, we embarked on the development of a new activity diagnostic based on infrared (\textit{WISE}), optical (SDSS) photometry, and machine learning methods. More specifically, this new diagnostic utilizes three colors in order to classify galaxies into five different activity classes: star-forming (SF), AGN, LINER, composite, and passive.

The paper is organized as follows. In Sec. \ref{data} we describe the data, introduce the photometry scheme and we describe the methods used for the selection of each galactic activity class. In Sec. \ref{sec3} we introduce the classification method. In Sec. \ref{sec4} we present the results of the training of our diagnostic tool and investigate its performance. In Sec. \ref{disc} we discuss the achieved results, the limitations of the tool and we explore the reliability of the classifier. In the same section, we compare our results with other widely used infrared classification methods for AGN. In Sec. \ref{concl} we summarize our conclusions.   
  
\section{Data accumulation} \label{data}
\subsection{The Sloan-Digital Sky Survey (SDSS)}

Our main galaxy sample is drawn from the \textit{Sloan Digital Sky Survey} (SDSS), a northern sky survey that provides homogeneous and high-quality photometric and spectral data. For the activity classification of the galaxies in our sample (see Sec. \ref{ac_cl}) we use the spectroscopic information provided by the SDSS-MPA-JHU catalog \citep{2003MNRAS.346.1055K,2004MNRAS.351.1151B,2004ApJ...613..898T}. This catalog includes spectroscopic line and redshift measurements for more than one million galaxies within the SDSS footprint. In order to obtain photometric data for these galaxies we crossed-matched the galaxies with reliable measurements (RELIABLE $\neq$ 0) with the SDSS - DR16 photometric catalog based on their specObjID. The SDSS - DR16 provides measurements for a number of surface brightness profiles and aperture sizes in five filters: \textit{u}, \textit{g}, \textit{r}, \textit{i}, and \textit{z}. For our purposes, we opted to use the fiberMag (flux within an appropriate to the SDSS spectrograph 3$^{\prime \prime}$ aperture) and the cModelMag photometry which is based on a radial profile which is a linear combination of the best fit of an exponential and a de Vaucouleurs profile. The former is a good approximation of the flux in a galaxy’s nucleus (especially for the nearby galaxies) and the latter gives the total flux of a galaxy in a given band. 

\subsection{The Wide-field Infrared Survey Explorer (WISE) photometry}

The \textit{Wide-field Infrared Survey Explorer} \citep[\textit{WISE};][]{2010AJ....140.1868W}, is a satellite that mapped almost the entire sky. The \textit{WISE All-Sky Release Source Catalog} has covered 42,195 $\deg^2$, or 99.86 \% of the entire sky in four broad bands in the $\sim$3–25 $\mu$m range. Its bands W1, W2, W3, and W4 have effective wavelengths at 3.4 $\mu$m, 4.6 $\mu$m, 12 $\mu$m, and 22 $\mu$m respectively. Their angular resolution was 6.1, 6.4, 6.5, and 12 arcseconds respectively. The \textit{WISE} survey provides several advantages for the classification of large populations of galaxies: it is more
sensitive than previous broad-band IR surveys; it covers the 3-25 $\mu$m range which includes several important diagnostic features, e.g. the polycyclic aromatic hydrocarbons (PAHs) emission features, primarily found in star-forming galaxies; and the 3-20 $\mu$m continuum probes the transition from the stellar continuum to dust emission of a galaxy that hosts an AGN.

The \textit{WISE} survey offers different photometry profiles and apertures. In this project, we will use the w?mag$\_$2 and w?gmag (the ? corresponds to different band numbers 1, 2, 3, and 4 for W1, W2, W3, and W4 respectively). The w?mag\_2 photometry is the calibrated source brightness measured within a circular aperture of 8.25 arcseconds radius centered on the source position for every \textit{WISE} band and no curve growth correction has been applied. The background sky was measured from an annulus with an inner and outer radius of 50 and 70 arcseconds respectively. The w?gmag photometry is based on elliptical aperture photometry for every \textit{WISE} band (the ? corresponds to 1, 2, 3, and 4 for W1, W2, W3, and W4 respectively). The parameters of the elliptical apertures (semi-major axis and position angle) are based on the 2MASS survey \citep{2006AJ....131.1163S}. In addition, the \textit{WISE} survey provides extended source photometry, which however, is subjected to significant photometric uncertainties due to the low signal-to-noise ratio in the lower-surface brightness regions of the galaxies.

As this project aims to study galaxies in the local Universe, we start our analysis by considering photometry from the w?gmag photometric aperture as these galaxies will appear extended in the \textit{WISE} apertures. The use of a fixed photometric aperture will result in missing some of the galaxy emission for the nearest galaxies and most importantly the inclusion of an increasingly larger galactic region for more distant galaxies. Although this may dilute some of the nuclear (AGN) emission, it allows the application of the diagnostic to a wide range of distances, from local galaxies to more distant unresolved ones. We find that $\sim$20\% of the galaxies in our sample appear extended in the \textit{WISE} apertures (\texttt{ext\_flg$\neq$0)}. This reduces aperture effects and allows the application of the diagnostic even to very local galaxies ($z \sim 0$).

For more distant objects that are unresolved by \textit{WISE} we use the w?mag\_2 photometry aperture. The reason for choosing the w?mag\_2 against other similar \textit{WISE} photometry apertures (e.g., w?mag\_1) was that the former has an aperture radius similar to that of the PSF. For each of the four individual \textit{WISE} bands, the w?gmag photometry is kept for all galaxies that have measurements in that aperture, and the w?mag\_2 photometry is used for all galaxies that do not have measurements on the w?gmag aperture. The consideration of the integrated photometry eliminates any aperture-related bias resulting from the large distance range of our galaxies, since galaxies that belong to the same activity class but have different distances will now have the same colors. Given the different photometry apertures available in the \textit{WISE} catalog, in order to overcome this photometric bias, our photometry consists of the extended apertures for the resolved and of the point-like apertures for the unresolved sources. In addition, spiral galaxies tend to have \ion{H}{II} regions scattered across the galaxy disk and a bulge region dominated by old stellar populations in the center \citep[e.g.,][]{2000A&A...355..949F,2001A&A...376..878O}. This hybrid photometry scheme is ideal for accounting for the infrared emission of these gas regions and also avoids confusion in the classification process due to aperture effects.  

\begin{figure}[ht]
\begin{center}
\includegraphics[scale=0.74]{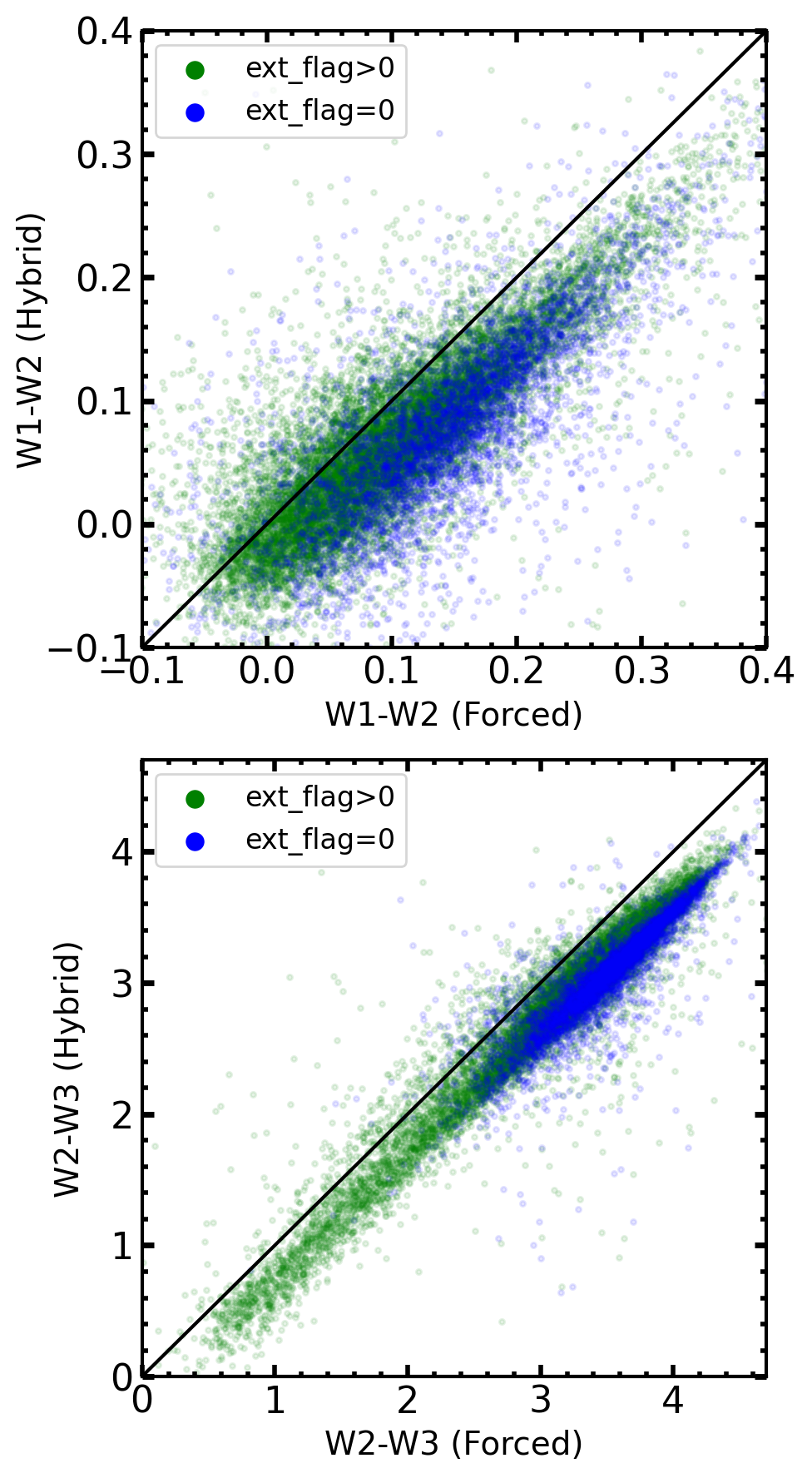}
\end{center}
\caption{Comparison plots between the forced photometry, WF, and the hybrid photometry scheme introduced in this work. On the top, the W1-W2 color calculated with the hybrid scheme against the same color but calculated the WF. On the bottom, we see the same plot for the W2-W3 color correspondingly. The galaxies have been color-coded according to their extension in the \textit{WISE} data (\texttt{ext\_flg=0} for point-like sources and \texttt{ext\_flg$\neq$0} for extended sources).} The black solid line is the $y=x$.
\label{fig:WF_vs_hyb}
\end{figure}

\cite{2016AJ....151...36L} provide \textit{WISE} integrated photometry by using higher-resolution \textit{WISE} maps together with apertures from the SDSS data (\textit{WISE} forced photometry or WF). This information is only available for the galaxies in the SDSS footprint and therefore not appropriate for an all-sky sample. Nonetheless in Fig. \ref{fig:WF_vs_hyb} we compare our hybrid photometry with the WF photometry. Since in our diagnostic we consider \textit{WISE} colors, in that plot, we see the one-to-one comparison of the two \textit{WISE} colors, W1-W2 and W2-W3, calculated with WF and with our hybrid scheme. This comparison shows that the two methods show good agreement apart from a small systematic offset of $\sim$0.1 mag in the W1-W2 color and $\sim$0.5 mag in the W2-W3 color. We also color-code the sources based on their \texttt{ext\_flg} value. If the source has \texttt{ext\_flg=0} means that its shape is consistent with a point-source profile in the \textit{WISE}. We see there is no dependency on the measured colors in our hybrid scheme depending on the source extent.

\subsection{Activity classes and passive galaxies} \label{ac_cl}

For the activity classification of galaxies with emission line spectra, we use the diagnostic tool defined by \cite{2019MNRAS.485.1085S}. This is an extension of the generally used diagnostics of \cite{1981PASP...93....5B,2001ApJ...556..121K,2003MNRAS.346.1055K,2007MNRAS.382.1415S}, that allows the simultaneous use of all available diagnostic line ratios avoiding contradictory classifications and providing more robust results. This scheme was based on fitting multivariate Gaussian distributions to the four-dimensional emission-line ratio distributions of $\log_{10}$([\ion{N}{II}]/H$\alpha$), $\log_{10}$([\ion{S}{II}]/H$\alpha$),
$\log_{10}$([\ion{O}{I}]/H$\alpha$), and $\log_{10}$([\ion{O}{III}]/H$\beta$). For this reason if we project these objects on the two-dimensional BPT diagram the tails of the distributions of the different activity classes may not be confined within the demarcation lines that separate the different activity classes defined by \cite{2001ApJ...556..121K,2003MNRAS.346.1055K,2007MNRAS.382.1415S}. The emission line measurements were obtained from the SDSS JHU-MPA catalog \citep{2003MNRAS.346.1055K,2004MNRAS.351.1151B,2004ApJ...613..898T}. The classes of galaxies considered in that classification scheme were: star-forming, Seyfert, LINER, and composite. This diagnostic was based on a probabilistic classifier. That means that based on the location of an object in this four-dimensional space, one can also determine the probability that it belongs to each one of the classes that the classifier has been designed to discriminate. In our analysis, we will use their Soft Data-Driven Analysis (SoDDA) classifier adopting the class with the highest probability. 

\begin{figure}[h]
\begin{center}
\includegraphics[width=3.5in,keepaspectratio]{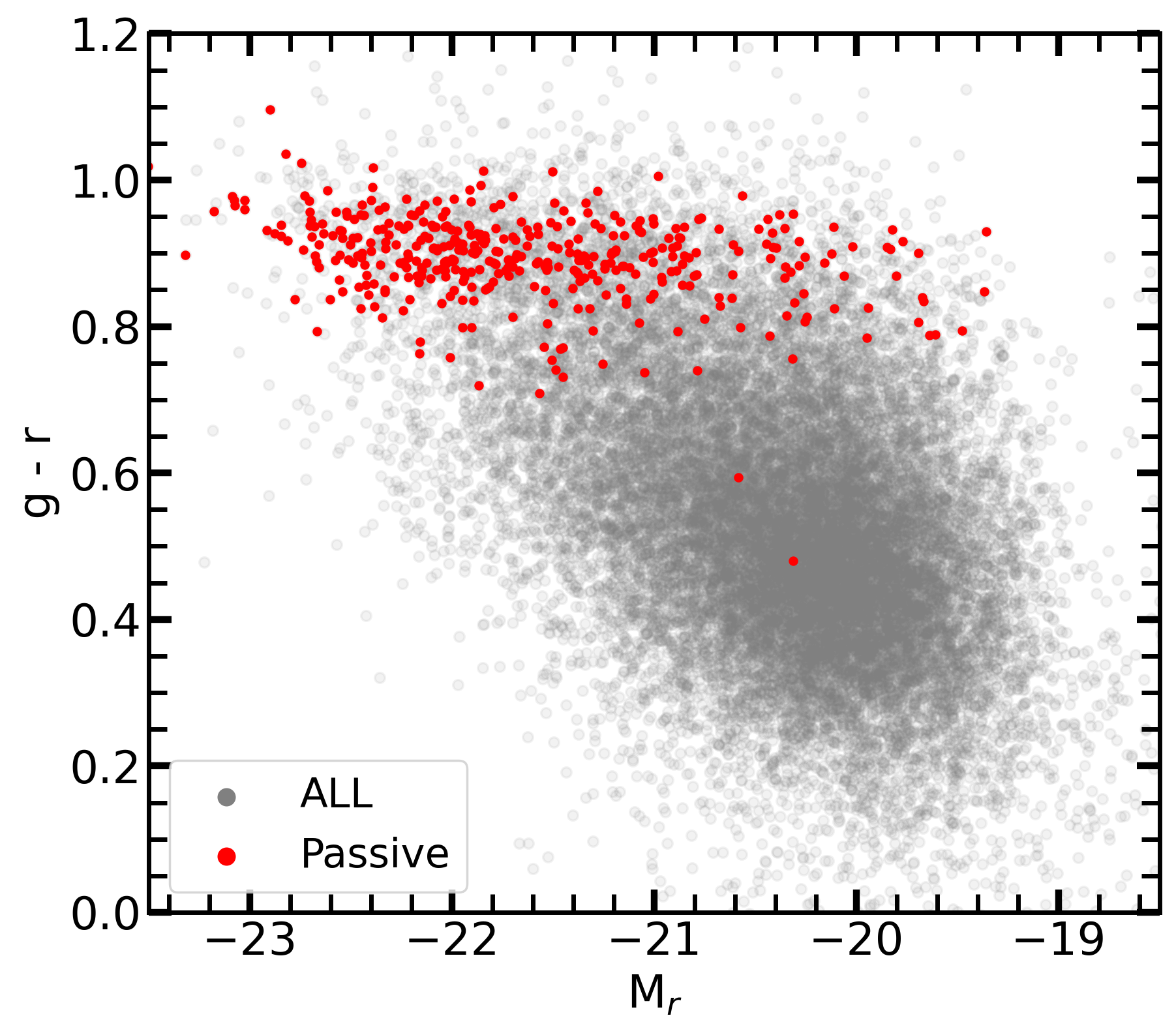}
\end{center}
\caption{A color-magnitude diagram (CMD) of \textit{g}-\textit{r} against M$_{r}$. On the y-axis is the \textit{g}-\textit{r} color against the absolute magnitude in SDSS r-band, M$_{r}$. The red points represent the sample of passive galaxies while the gray points represent the whole sample of galaxies (all classes).}
\label{fig:CMD_passive}
\end{figure}

So far, most galactic emission-line diagnostic tools do not include the class of passive galaxies. Inactive or passive galaxies are defined as galaxies that do not show any evidence of activity (i.e., star formation or AGN) based on the lack of optical emission lines. Since in this work, we also consider the class of passive galaxies, we need to define the corresponding sample. Thus, since we seek non-active galaxies the sample of passive galaxies was selected by the following criteria: emission lines of H$\alpha$, H$\beta$, [\ion{O}{III}] $ \lambda$5007, [\ion{O}{I}] $\lambda$6300, [\ion{N}{II}] $\lambda$6584, and [\ion{S}{II}] $\lambda \lambda$6717,6731 should have signal-to-noise ratio below 3 and the signal-to-noise ratio of the continuum at the location of each emission line is above 3. This ensures that the lack of emission lines is not the result of the difficulty in measuring them in poor-quality spectra. A confirmation that this method of classifying galaxies as passive is effective, is their location on the color-magnitude diagram \citep[e.g.,][]{2004ApJ...608..752B}. Figure \ref{fig:CMD_passive} shows the \textit{g}-\textit{r} color against the absolute r-band photometry (M$_{r}$). The galaxies selected spectroscopically as passive are located on the upper part of the diagram in the so-called red sequence region, where early-type galaxies are found. 

\subsection{Final sample} \label{sec24}

After defining the criteria for the selection of each galaxy class, we filter the galaxies that will constitute the final training sample based on the quality of the photometric data and the activity classification.

As our goal here is to train a machine learning algorithm, the filtering has to be done in two stages. The first one will ensure that the true labels (i.e., activity classes) are well defined, as a poor true label definition based on insecure classification can lead to an algorithm with significant uncertainty in its predictions. The other stage regards the features (photometric measurements) that will be used for the discrimination between the different galaxy classes by the new activity diagnostic tool. 

For the first step, we only select active galaxies that have S/N above 5 for all optical emission lines that were used for the characterization of the true class of each galaxy (Sec. \ref{ac_cl}), namely, H$\alpha$, H$\beta$, [\ion{O}{III}] $ \lambda$5007, [\ion{O}{I}] $\lambda$6300, [\ion{N}{II}] $\lambda$6584, and [\ion{S}{II}] $\lambda \lambda$6717,6731. As stated earlier, the classification of each galaxy is based on the class with the highest probability in the diagnostic of \cite{2019MNRAS.485.1085S}. As the classifier also provides probabilities for each galaxy to belong in the other considered classes, we choose galaxies that have been classified with high confidence based on the probability difference of the highest and second higher predicted probability that were assigned by the classifier for each galaxy. Classifications with a large difference between the first and the second-ranking class are considered highly reliable. In this respect, we consider galaxies with a difference in their predicted probabilities of at least 25\%.

First, we consider objects with reliable photometric measurements based on the \textit{WISE} quality flags. To identify and remove these problematic cases we consult the \textit{AllWISE Source Catalog and Reject Table} \footnote[1]{\url{https://wise2.ipac.caltech.edu/docs/release/allwise/expsup/sec2_1a.html}}, where we find the quality flags for the photometry of a galaxy in the first three \textit{WISE} bands. We consider as unreliable photometry every detection that, in at least one of the three \textit{WISE} bands (1, 2, and 3), has been flagged in the above-mentioned catalog as having a measurement error of $\texttt{9.999}$, as this indicates that even though a measurement exists it should be considered as highly suspicious. Also, another flag concerning the quality of the photometry is the $\texttt{w?flg=32}$. Every galaxy with this flag means that its photometry measurement is in the 95\% upper limit and should not be considered reliable detection. Other important factors that have to be accounted for in the quality of the photometry measurements are source contamination and confusion. If a source has been flagged in $\texttt{cc\_flags}$ with a value of $\texttt{D, P, H, O, d, p, h, or o}$, it means that the source may be contaminated due to its proximity to an image artifact and thus we remove any galaxy that has one of these flags in any of the W1, W2, and W3 bands. Concerning the second stage of filtering, we choose active and passive galaxies that have photometric data with S/N > 5 in the two \textit{WISE} bands, W1 and W2, as well as for the two SDSS filters (\textit{g} and \textit{r}). A more relaxed lower limit of S/N = 3 for the W3 \textit{WISE} band is selected. The reasoning behind these choices is that the W3 \textit{WISE} band has lower sensitivity than the W1 and W2 bands and a strict S/N selection criterion will result in a significant reduction in the number of galaxies in our sample.

Other important facts that have to be taken into consideration are survey selection and galaxy evolution effects. We find that in our sample the number of AGN and passive galaxies tend to increase sharply with redshift. In order to create a uniform distribution of activity classes of galaxies across the whole redshift range, we split the sample into four equal redshift bins. Our sample of galaxies spans the redshift range from $z= 0.02$ to $z= 0.08$. We limit the lower cutoff of redshift to $z= 0.02$ as this is compatible with the definition range of the BPT diagrams (i.e., from $z \sim$ 0.02 to $z \sim$ 0.06) and thus avoids strong aperture effects during the training of the algorithm. We proceed by finding a "reference" redshift bin which will be used as the basis for selecting the number of objects to sample from each class in each redshift bin. We find that the $0.033 < z < 0.047$ bin is ideal as a reference bin as it is close to the middle of the redshift range. Based on this bin we randomly select the same number of objects for each class individually from the other three redshift bins (namely $0.02 < z < 0.033$, $0.047 < z < 0.063$, and $0.063 < z < 0.08$.) 

After the implementation of the two stages of filtering, the redshift balancing, and the removal of unreliable detections in the \textit{WISE} photometry we obtain the final sample that contains all the eligible galaxies for the training process of our diagnostic tool. In that sample, there are 40954 galaxies in total, with redshifts between $z=0.02$ and $z = 0.08$. The composition of the training sample per galactic activity class is given in Table \ref{tab:nonlin}. We note that although our classifier is trained on a sample with high quality optical spectroscopy classifications, it can be used on any sample of galaxies with available photometry in the \textit{WISE} and SDSS bands. The only limitation is that the IR photometry should encompass the extent of the galaxy, and the optical photometry the central 3$^{\prime \prime}$ of the galaxy (to match the SDSS fiberMag).

\section{The diagnostic tool} \label{sec3}
\subsection{The random forest algorithm} \label{secrf}
 
For the development of our diagnostic we opted to use the random forest algorithm \citep{2014arXiv1407.7502L}, which is based on the concept of decision trees. A decision tree starts with a root node that contains all the training data, then it will use the considered features to progressively create more homogeneous groups of data (nodes). Ideally, at the end of the process, the final nodes (leaves) will only contain data of the same kind (class). The problem with a single decision tree is that, in most cases, the tree tends to adapt too well to the training data, and as a result, its performance is poor when it is applied to new data (overfitting). To avoid overfitting, we can combine many decision trees in parallel to build a random forest. Each decision tree of the random forest is trained on a subsample of the training data. Every such subsample of the full dataset that is used for the training of the trees is selected by randomly shuffling the full training set. 

During the classification process, each tree takes as input an object and gives as output (or vote) the class that this individual object belongs. Then, this process continues until that object has been through every tree of the ensemble. In the end, the decisions made by every tree of the ensemble for the object under question are summed and the object belongs to the class that collected the most votes. The algorithm also allows us to calculate the probability of that object belonging to each one of the classes. This probability is given by the ratio of the number of votes the object received to belong in a particular class to the total number of trees considered in the algorithm.

It is called random because, during the training process of the algorithm, the features used to make the split of the data into the new nodes are selected randomly. The random forest offers several advantages: it is intuitive, probabilistic and it is easily adaptable to many problems. We use the implementation of the
\begin{figure}[H]
\begin{center}
\includegraphics[scale=0.7]{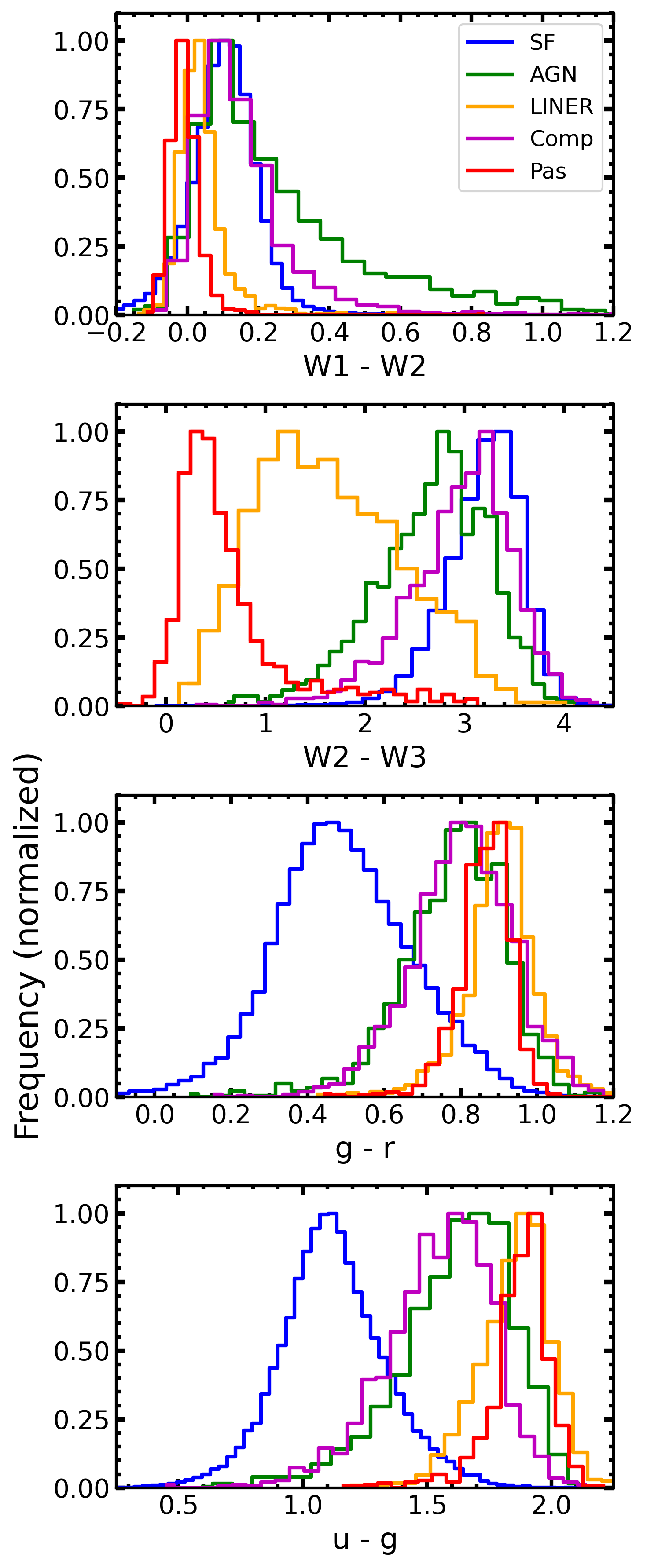}
\end{center}
\caption{Distributions of colors considered as potential features for the definition of our diagnostic tool. Starting from top to bottom, we see the distributions of colors W1-W2, W2-W3, \textit{g}-\textit{r}, and \textit{u}-\textit{g} for each galactic activity class. Combinations of these four colors will be used as potential feature schemes for defining our diagnostic. Due to the high imbalance of the sample, the number of galaxies is normalized based on the frequency of occurrence in our data sample. Blue histograms correspond to the starforming (SF), green to the AGNs, yellow to LINERs, purple to composite, and red to the passive galaxies.}
\label{fig:ft_dist}
\end{figure}

\begin{table}[ht]
\caption{Composition of the final sample per galactic activity class.} 
\centering 
\begin{tabular}{l c c}
\hline\hline
  Class & Number of objects & Percentage (\%)\\
\hline
Star forming & 35878 & 87.6\\
Seyfert & 1337 & 3.3\\
LINER & 1322 & 3.2\\
Composite & 1673 & 4.1\\
Passive & 744 & 1.8\\
\hline
\end{tabular}
\label{tab:nonlin} 
\end{table}

\noindent random forest algorithm provided by \texttt{sklearn.ensemble.RandomForestClassifier()} from the \texttt{scikit-learn} Python 3 package, version 1.1.2.

\subsection{Performance metrics} \label{sc33}

\begingroup
\setlength{\tabcolsep}{10pt} 
\renewcommand{\arraystretch}{1.5}
\begin{table*}[ht]
\caption{Definition of each performance metric used for the evaluation of the performance of our diagnostic tool.}
\centering
\begin{tabular}{p{0.2\linewidth}p{0.5\linewidth} c}
\hline\hline
 & Description & Equation\\
\hline
Term & & \\
\hline
True Positive (TP) & An object that has been correctly predicted to be in a specific galaxy class. & -\\
True Negative (TN) & An object is correctly predicted not to belong to a class. & -\\
False Positive (FP) & An object that is falsely predicted to belong to a galaxy class. & -\\
False Negative (FN) & An object that is falsely predicted not to belong to a galaxy class. & -\\
\hline
Performance metric & & \\
\hline
Accuracy & The ratio of the correct predictions to the total predictions made by the classifier. & $\frac{TP+TN}{TP+TN+FN+FP}$ \\
Precision & The number of objects correctly predicted to belong to a class divided by the total objects that the classifier predicted to belong to that class. & $\frac{TP}{TP+FP}$\\
Recall & The objects correctly predicted to belong to a class divided by all the objects that belong to that specific class. & $\frac{TP}{TP+FN}$\\
F1-score & The harmonic mean of precision and recall. & $\frac{2TP}{2TP+FP+FN}$ \\
\hline
\end{tabular}
\label{table2}
\end{table*}
\endgroup

To evaluate the performance of our diagnostic tool we adopt standard metrics such as the accuracy, the precision, the recall, and the F1-score. The exact definition of each performance metric we use for the evaluation is presented in Table \ref{table2}. When evaluating the performance of an algorithm based on the accuracy metric may lead to misleading results, especially in cases with skewed data sets like the one we are dealing with here. 

The metrics that we consider as more appropriate are the recall and the precision for each class. Recall is a metric that quantifies completeness (how many objects of each class have been correctly selected) while the precision quantifies contamination (the fraction of correctly selected objects within the population of all selected elements). For the proper evaluation of the performance, we plot the confusion matrix and we calculate the precision and the recall scores per class.

\subsection{Feature selection} \label{sec34}
 
There are many characteristics that one can use to classify galaxies based on their activity. Our main goal here is to define a diagnostic tool that is capable of discriminating efficiently between the galaxy classes by utilizing observables that can easily be acquired for a large number of galaxies. Considering all the above, in this work, we use only infrared and optical colors that are available from all-sky or wide-area surveys. Initially, we start by considering colors that are the combinations of the three \textit{WISE} bands and three SDSS filters. More specifically using combinations of the \textit{WISE} bands 1, 2, and 3 along with the \textit{u}, \textit{g}, and, \textit{r} SDSS (fiberMag photometry) filters we calculate the colors: W1-W2, W2-W3, \textit{g}-\textit{r}, and \textit{u}-\textit{g}. In Fig. \ref{fig:ft_dist} we see the distributions of each color we consider as a potential feature, for the different classes, normalized by the total number of objects in each activity class. In our analysis, we also considered including the W3-W4 color as a potential feature. However, due to the low sensitivity of the detector at the 24 $\mu m$ band, combined with the weak emission from passive galaxies in the mid-infrared, we found that almost none of the passive galaxies has reliable detections in the W4 \textit{WISE} band. Therefore, this band (and the colors involving it) is not considered further in our study. 

These particular features have been chosen based on the broad-band spectral shape of these five different galaxy classes. Previous diagnostics have demonstrated the diagnostic power of \textit{WISE}. Star-forming galaxies have \ion{H}{II} regions which are rich in dust and gas heated up by hot young stars, producing strong emission in the infrared \textit{WISE} bands (in particular in W3 due to the PAHs and dust). The AGN hosting galaxies have a rising red continuum due to dust heated by the power-law UV spectrum while at the same time, this extreme UV radiation environment results in the dissociation of the large molecules, leading to suppressed PAH emission \citep{2014MNRAS.443.2766A}. However, the two top plots of Fig. \ref{fig:ft_dist} show that there is \text{an} overlap between the classes in the \textit{WISE} colors, especially in the case of the W1-W2 color. In the two bottom plots of the same figure, we see that the optical colors help to break the degeneracy observed in the mid-IR colors.

\begin{figure}[ht]
\begin{center}
\includegraphics[width=3.5in,keepaspectratio]{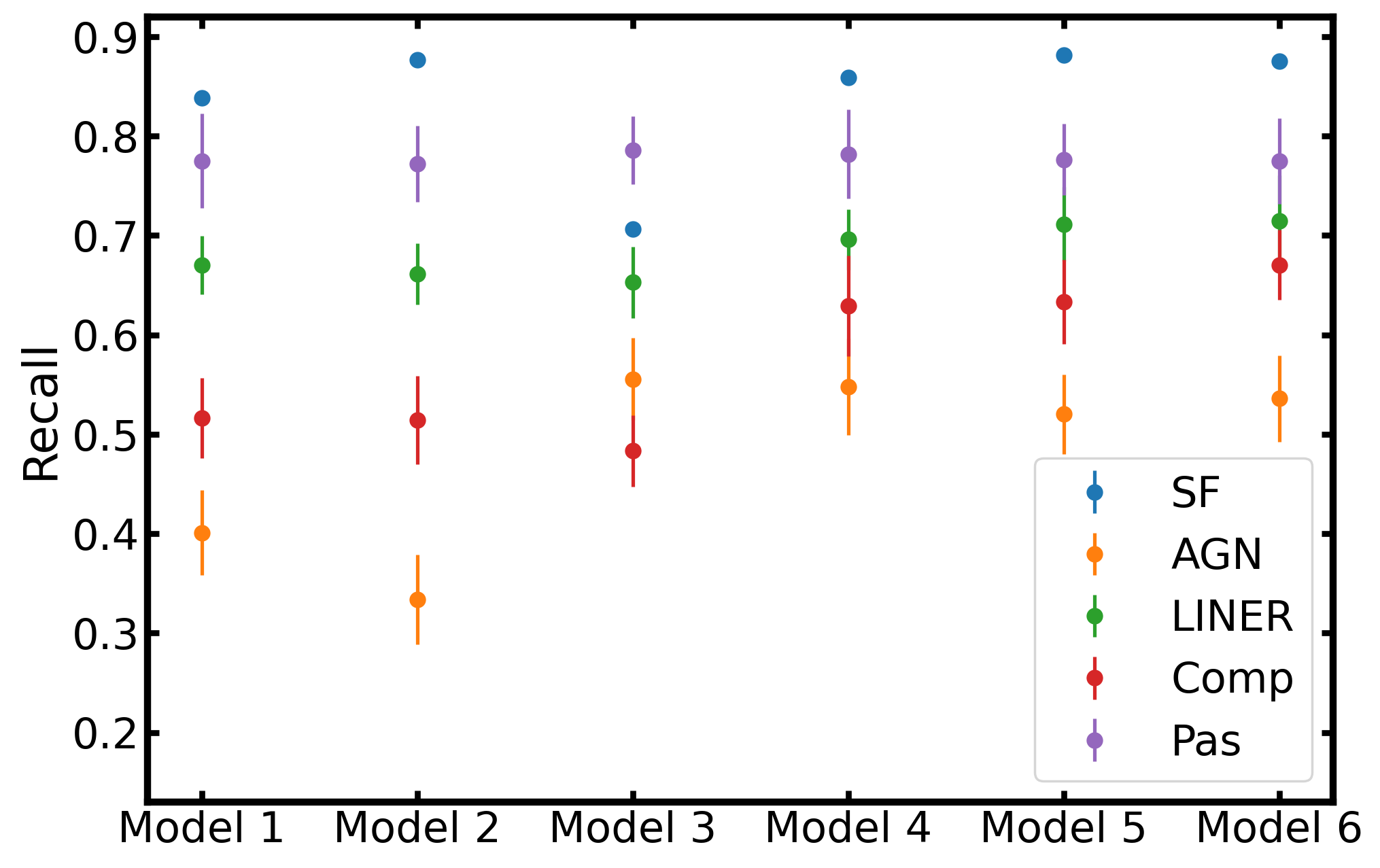}
\end{center}
\caption{Recall scores of the different models (features schemes) considered for our diagnostic tool. The description of the models is presented
in Table \ref{table3}. The error is the standard deviation of the k recall
scores (here k = 10), calculated with the cross-validation method. The error on the recall scores for the SF galaxies is too small to be depicted here. Blue points correspond to the starforming (SF), yellow to the AGNs, green to LINERs, red to the composite, and purple to passive galaxies.}
\label{fig:rec_models}
\end{figure}

Even though there is astrophysical reasoning behind our initial feature selection, we can not know a priory which combinations give optimal results and which ones do not provide any improvements in the performance of this multi-class classification problem. In order to determine which combination can yield optimum results with the minimum number of features we train the algorithm with different combinations of features and we record the performance of each training scheme. This method helps us to compare the performance of the different models and also helps us to identify highly informative or redundant features. The total number of models examined is six and they are presented in Table \ref{table3}. In Fig. \ref{fig:rec_models} we plot the recall score of each activity class for each model (i.e., feature scheme) presented in Table \ref{table3}. We start from the simplest case by testing one infrared and one optical color. For the first two simple models from the two available infrared colors, we decide to test only the W2-W3 combined with a different optical color each time as the W2-W3 provides greater dynamic range and discrimination between the different classes compared to the W1-W2. The selection of a feature scheme (model) for this new diagnostic will be made on two basic criteria: (1) it has to offer high recall scores for each class and (2) it uses the minimum number of features. We chose to use the recall as a performance metric since it also offers information about completeness.

\begin{table}[ht]
\caption{Different combinations of features (colors) that were tested as potential models for the definition of the diagnostic.} 
\centering 
\begin{tabular}{l c}
\hline\hline
  Scheme & Features (colors) \\
\hline
Model 1  & W2-W3, \textit{g}-\textit{r} \\
Model 2  & W2-W3, \textit{u}-\textit{g} \\
Model 3  & W1-W2, W2-W3 \\
Model 4  & W1-W2, W2-W3, \textit{g}-\textit{r} \\
Model 5  & W1-W2, W2-W3, \textit{u}-\textit{g} \\
Model 6  & W1-W2, W2-W3, \textit{g}-\textit{r}, \textit{u}-\textit{g} \\
\hline
\end{tabular}
\label{table3}
\end{table}

In our analysis we use the average value of the recall scores calculated with the cross-validation method. In more detail, we use galaxies from the final sample (see Sec. \ref{sec24}) by imposing the additional criterion that all galaxies must have S/N > 5 in the u SDSS band. Then, we split those galaxies in k-folds and we use the k-1 folds for training and one for the performance evaluation of the model (testing fold). Every time, we replace the testing fold with one of the training folds and we repeat until all the k folds have been in the position of the testing fold. We keep the k total recall scores and calculate the average. The error bars on the average recall are the standard deviation of the k recall scores. We select the number of folds to be 10 (k = 10) which is the maximum number of folds that offers a balance between satisfactory number of objects in each fold for the under-represented classes and good statistics for evaluating the performance metrics. 

After evaluating each possible model, we conclude that the model which includes all the available colors (Model 6) does not improve the performance when compared to the models which use two infrared and one of the two optical colors (Model 4 and 5). The performance of Model 3 is generally the same as that of Models 4, 5, and 6, but the lack of an optical color results in dramatically lower recall for the star-forming galaxies. This can be attributed to the fact that star-forming galaxies have bluer \textit{g}-\textit{r} colors which separate them clearly from the other activity classes (Fig. \ref{fig:ft_dist}). For the other two models (Models 1 and 2) we notice that the classifier does not have enough information to separate the classes effectively. In particular, we notice that there is a significant performance drop for the AGN and composite galaxy classes. This can probably be explained by the nature of the AGN and composite galaxies. The W2-W3 color records the infrared emission from the circumnuclear dust heated by the supermassive black hole which is found in every AGN galaxy but can also be present in a composite galaxy. Although AGN-heated dust can lead to stronger emission in the W2 band, the majority of the AGN-hosting galaxies have similar W1-W2 colors to star-forming and of course to composite galaxies. This creates confusion on a classifier that is defined based only on the feature schemes of Models 1 and 2 leading to an extensive mixing between these two classes. After these results, we decide to adopt Model 4 as our basic model, which has the following features: two infrared colors (\textit{WISE}), W1-W2, W2-W3, and one optical color (SDSS), the \textit{g}-\textit{r}. In Fig. \ref{fig:rec_models}, we see that Model 5 has a similar performance as Model 4. However, Model 5, relies on u-band photometry which often has lower signal-to-noise measurements than the r-band, limiting the applicability of the classifier to larger data samples. 

After the optimal combination of features has been determined we proceed with the optimization of the algorithm. In this process we search for the values of the algorithm's hyperparameters that offer the best performance in a particular problem. Upon investigation, we find that by tweaking nearly half of the algorithm's hyperparameters the scores do not improve significantly and thus they are left in their default values, as imported from \texttt{scikit-learn}. 
The hyperparameters that have a significant impact and hence are worth optimizing are the following: \texttt{max\_depth, max\_leaf\_nodes,
max\_samples, min\_samples\_leaf, min\_samples\_split}, and \texttt{n\_estimators}. The exact procedure of the optimization as well as a table (Table \ref{tablehyp}) with the best values for each important hyperparameter are presented in the appendix \ref{fig:val_crvs}.

\subsection{Implementation} \label{sec32}

The standard procedure for this step is to separate the sample of all available data, into three random subsets. One of them contains the majority of the objects (50\% of the total or 20476 galaxies) and it will be used for the training. The data from that subset are used to adapt the algorithm to the individual problem. The rest of the data form the test and the validation set. The validation is used for the calibration of the classifier while the test set is only used for the evaluation of its performance after the training and calibration processes (see Sec. \ref{calcl}). For this project, we perform a training-validation-test set with proportions of 50\%-25\%-25\% or 20476-10239-10239 galaxies respectively. The split is stratified which ensures that each subset has the same percentage of objects in each class as the original sample.

The high imbalance in the number of objects between the five classes in the training sample cannot be left unnoticed, since it can lead to biases in the classification in favor of the class that has the higher frequency of appearance. To avoid such an effect we have two options: one is to reduce the sample in such a way that every individual class has the same number of objects and the second is to assign weights to each class, i.e. to adjust the impact of each object will have during the training of the algorithm. In the implementation of the random forest we use here, these weights are calculated internally by setting the \texttt{class\_weight} parameter to "balanced\_subsample". We select the second option, as it makes use of all available data leading to a more robust classification tool as the training will contain a broader range of examples.

\subsection{Classifier calibration} \label{calcl}

Every probabilistic classifier provides us with not only the class of the object under investigation but also the probability that this object belongs to each one of the classes that the classifier was designed to discriminate. When the classes are very well separated in the feature space these probabilities represent the actual likelihood of finding an object to belong to a certain class given its location in the feature space. However, in classification problems where there is some mixing between the distribution of the features for the considered classes, these probabilities do not necessarily reflect the actual probability of the true class. To correct for this effect in our classification problem we calibrate our classifier. This way the output of the classifier represents the actual probability of finding an object of a particular class within a given region of the feature space. 

In the case of a multi-class classification problem, the process of calibration is performed in a One-vs-Rest fashion, i.e. following the same process as in the binary classification for each class individually, and splitting the multi-class problem into multiple binary-class problems. The calibration in a binary classification is achieved by applying a regression algorithm (calibrator) that rescales the raw predicted probabilities so that these probabilities match the expected distribution of the actual probabilities. The latter is based on the frequency of an object of a given class appearing among the sample in the feature space. The subsample that is used for the calibration must contain objects that have not been used during the training of the classifier to avoid any bias.

The data we use for the calibration of our diagnostic are from the validation subset (see Sec. \ref{sec32}), a held-out subset of the data that was not a part of the training process. The algorithm we use to perform the calibration is the \texttt{CalibratedClassifierCV} which is provided by the \texttt{scikit-learn} package. We opt to use the "sigmoid" over the "isotonic regression" method as the latter is prone to overfitting, especially in problems with severely under-presented classes. Due to the fact that star-forming galaxies are the overwhelming majority of the objects in our sample, a feature that is preserved in the stratified split of the data among the three subsets of data (i.e., training, validation, and test set), the validation set has an excess number of star-forming galaxies.

This imbalance in the number of objects in each class can lead to biases in the calibration in favor of the class with the higher frequency of appearance. To avoid such effects, only for the validation set, we manually balance the sample by keeping the same number of objects from each class.

\section{Results} \label{sec4}

\subsection{Performance} \label{sec41}

\begin{figure}[h]
\begin{center}
\includegraphics[scale=0.32]{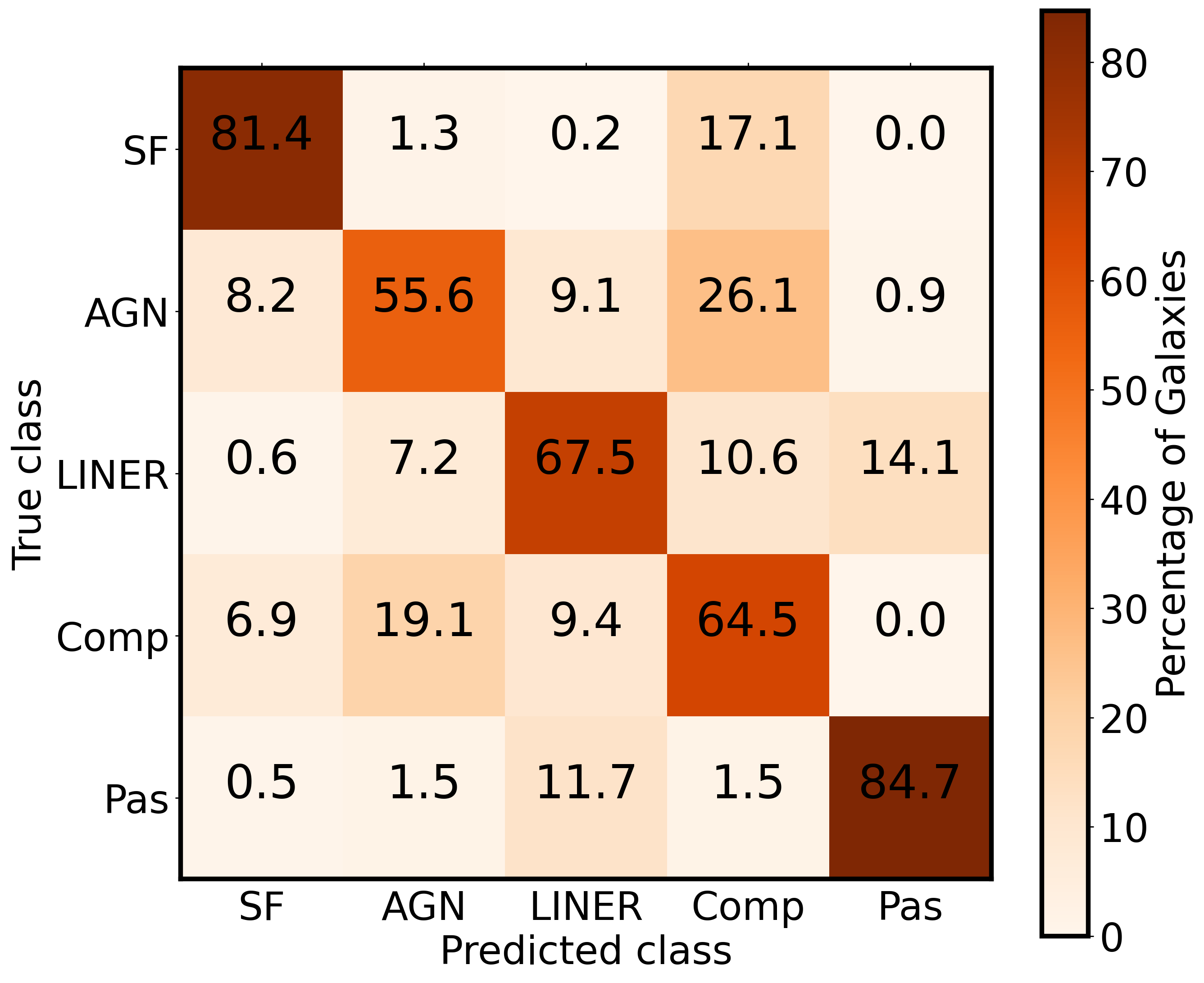}
\end{center}
\caption{Confusion matrix for the test subset of galaxies. The numbers (and the color-code) in this plot represent the percentage of the classified objects in each class with respect to the total true population in each class. The labels on the x- and y-axis represent the predicted and the true class of a galaxy respectively. Labels: SF, star-forming; Comp: composite; and Pas, passive.}
\label{fig:cnf_mt}
\end{figure}

We start by calculating the overall accuracy of our diagnostic. As before, we use the k-fold cross-validation method. However, here we use a reduced number of folds because, in this case, we have to keep separate an additional subset of data from the final sample which is going to be used for the probability calibration of the algorithm. We find that the maximum number of folds we can split our data maintaining enough objects to adequately represent the minority classes and still have good statistics is five. The overall accuracy achieved is 81\% $\pm$ 1\%. We acknowledge that the accuracy alone is not enough to describe the performance. Thus, we consider additional performance evaluation metrics, such as the confusion matrix. This is a two-way table in which the lines (y-axis) represent the true labels and the columns (x-axis) are the predicted labels for each data point made by the classifier. Each cell gives the fraction of the objects from a given true-class that is classified in each of the considered classes (columns). Therefore, the confusion matrix is the summary of the results made by the algorithm when evaluated on the test subset on a class-by-class basis. In a perfect classifier, all objects populate only the primary diagonal ($y=-x$). The confusion matrix provides information not only for the number of correct predictions but also about the objects that were misclassified, by checking what classes the classifier mixes. In Fig. \ref{fig:cnf_mt} we present the confusion matrix calculated on the test subset of the final data sample defined in Sec. \ref{sec24}. Inspecting the confusion matrix we conclude that the overall performance is good, with the higher scores achieved for the classes of the star-forming and passive galaxies. 

Based on these results we can also calculate additional metrics that can give us a more detailed view of the performance per class. In Table \ref{tab:cl_reprt} we present the values for the metrics defined in Table \ref{table2} calculated for each class. In that table we see that the star-forming and passive galaxies have excellent scores while the rest of the classes (AGN, LINER, and composites) have good to moderate performances.

\begin{table}[ht]
\caption{Performance metrics calculated for each class.}
\centering
\begin{tabular}{l c c c c }
\hline\hline
Class & Precision & Recall & F1-score & Number of \\ & & & &galaxies \\
\hline
Star forming & 0.99 & 0.81 & 0.89 & 8960\\
Seyfert & 0.45 & 0.56 & 0.50 & 329\\
LINER & 0.66 & 0.68 & 0.67 & 320\\
Composite & 0.14 & 0.65 & 0.24 & 434\\
Passive & 0.76 & 0.85 & 0.80 & 196\\
\hline
\end{tabular}
\label{tab:cl_reprt}
\end{table}

\subsection{Feature importance} \label{sec42}

A useful output of the random forest algorithm is the feature importance which describes the relevance of each feature during the training of the classifier. Therefore, it provides a measure of how much a given feature contributes to the ability of the random forest to discriminate between the different classes. So, a feature that clearly characterizes a class will have high relevance (or importance). Furthermore, it provides insights into the physical parameters that drive the performance of the classifier.

\begin{figure}[ht]
\begin{center}
\includegraphics[scale=0.7]{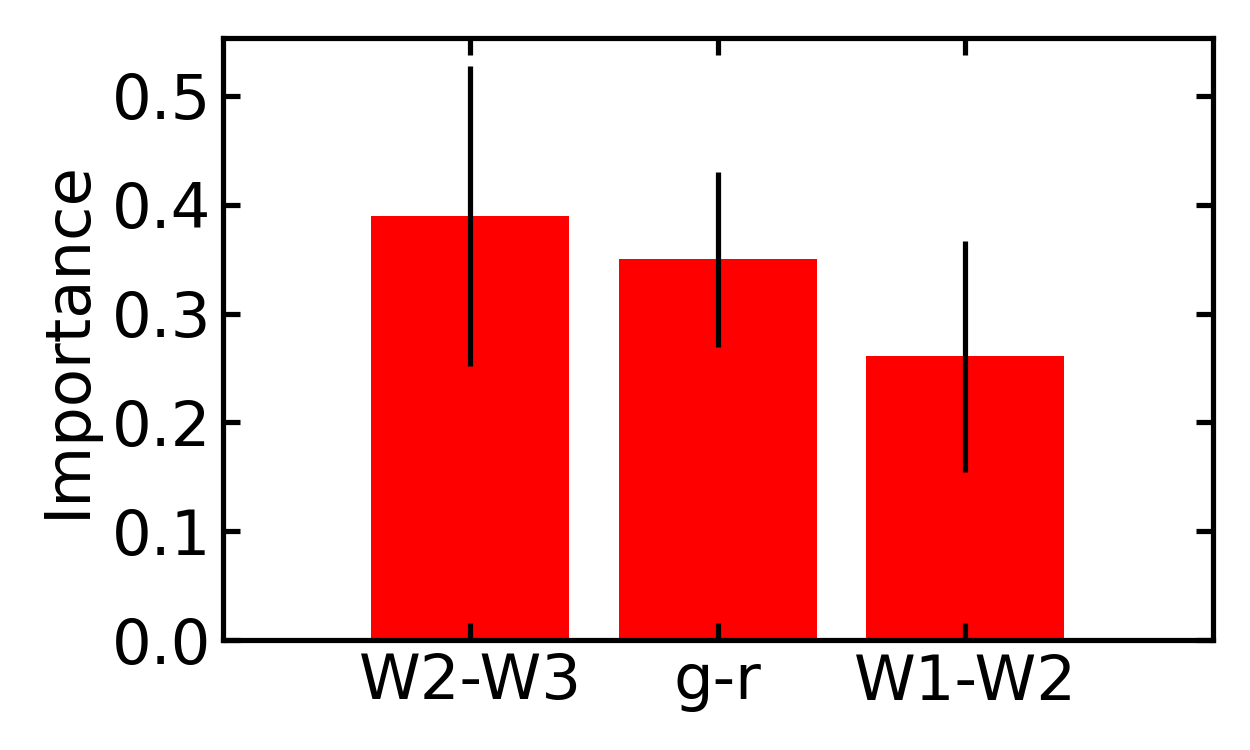}
\end{center}
\caption{The importance (relevance) of the three features used for the definition of the diagnosis as were calculated during the training of the random forest. The W2-W3 color is the most important feature, while all features are of comparable relevance.}
\label{fig:ft_imp}
\end{figure}

In Fig. \ref{fig:ft_imp} we present a bar plot of the feature importance scores. As the feature importance is calculated in each node, we can calculate the average and its standard deviation. From Fig. \ref{fig:ft_imp} we see that the chosen feature scheme is well-defined, as all features play a similar role in the classification of these five activity classes. Thus the feature importance helps us to better understand the operation of this algorithm.

\subsection{Application on the different redshift subsamples} 

\begin{figure}[h]
\begin{center}
\includegraphics[scale=0.42]{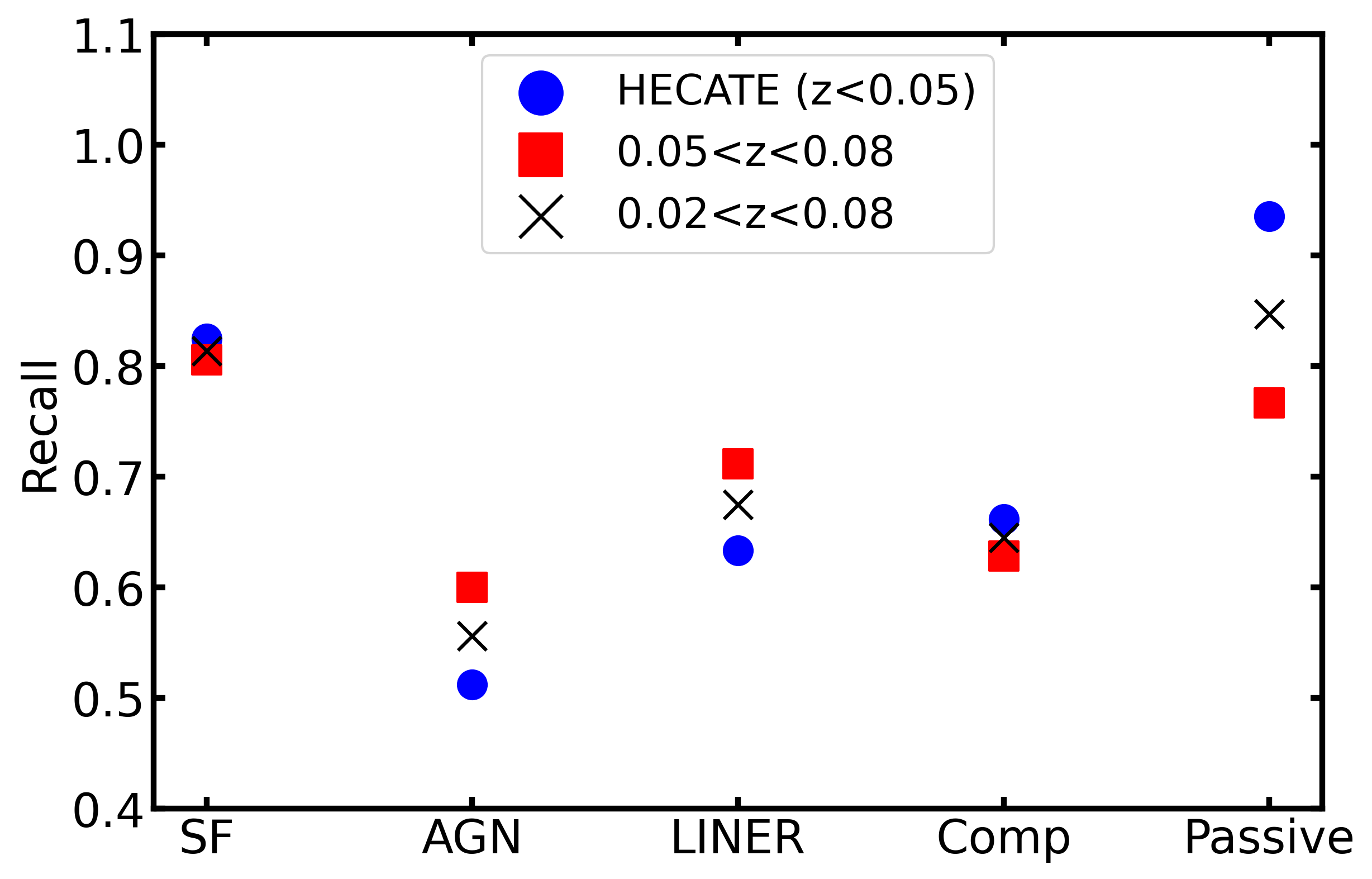}
\end{center}
\caption{The fraction of the correctly classified objects to the total true objects for each class, for three redshift ranges. The first redshift range is from $z=0.02$ to $z=0.05$ (galaxies in the HECATE catalog \citep{2021MNRAS.506.1896K} indicated by blue disks in the plot, the second redshift bin has ranged from $z=0.05$ to $z=0.08$ indicated by red squares. The third redshift range includes galaxies from the whole redshift range that the diagnostic was trained on, $0.02<z<0.08$, which is indicated by black x marks. Labels: SF, Star-forming; Comp, Composite.}
\label{fig:rcl_cl_z}
\end{figure}

After we have trained and optimized the diagnostic tool we proceed by applying it to two different subsets of the test set (Sec. \ref{calcl}), spanning two different redshift ranges: $0.02<z<0.05$ and $0.05<z<0.08$. Since the initial training of the algorithm was performed on the full redshift range of $z$ = 0.02 to $z$ = 0.08, this exercise shows whether the performance of our diagnostic has a redshift dependence. The number of true objects per class in the $0.02<z<0.05$ redshift range is as follows: 3982 SF, 164 AGN, 150 LINER, 216 composite, and 93 passive galaxies. In the redshift range of $0.05<z<0.08$ the number of objects is as follows: 4978 SF, 165 AGN, 170 LINER, 218 composite, and 103 passive galaxies. In Fig. \ref{fig:rcl_cl_z} we present the recall score for each class in these two redshift bins when we apply our diagnostic to each one separately. For reference, we also show the scores of the diagnostic in the overall redshift range (0.02 < $z$ < 0.08). In this figure we see that the diagnostic has similar behavior (similar recall scores) for star-forming galaxies for the whole redshift range of our sample of galaxies. Though, we notice that AGN have slightly reduced performance in the lower redshift compared to the higher redshift bin ($\sim$ 20\% lower recall compared to the bin having higher redshift galaxies). Similar discrepancy is seen in the case of LINER and Passive galaxies. The reason for this behavior is discussed in Sec. \ref{secm53}. 

\section{Discussion} \label{disc}

In this work, we defined a new all-inclusive (i.e., including active and passive galaxies) diagnostic tool based on the combination of commonly available mid-IR and optical colors. In the following sections, we further discuss the behavior and robustness of the diagnostic and we compare it with other commonly used diagnostics.

\subsection{Physical interpretation of the results}

So far we have seen that our diagnostic achieves very good performance despite the limited information it uses. The use of the optical color of the galactic nucleus seems to have an important role in the efficient classification of galaxies (see Figs. \ref{fig:rec_models} and \ref{fig:ft_imp}). In Fig. \ref{fig:ft_dist} we saw that the five activity classes present different behavior in terms of the distribution of the colors considered in our diagnostic.

In particular, in the case of star-forming and composite galaxies, we see higher values of the W2-W3 color attributed to significant emission from PAHs in the W3 band. On the other hand, passive galaxies are poor in dust and their stellar populations are older, resulting in a declining emission in redder wavelengths. In contrast, AGNs show rising emission in the mid-IR in all three \textit{WISE} bands. This can be explained by emission from the dusty torus around the accretion disk. However, in contrast to star-forming galaxies they have weak PAH emission since these sensitive molecules are destroyed by the strong UV radiation from the accretion disk \citep[e.g.,][]{2014MNRAS.443.2766A} or their emission is diluted by the AGN continuum \citep[e.g.,][]{1998Natur.391...17G}.

Composite galaxies have weaker continua in the 3-12 $\mu$m range than AGN, but with stronger PAH emission, which however is weaker than that of star-forming galaxies. They also show strong silicate absorption. This is reflected in their W1-W2 and W2-W3 colors which are intermediate to those of AGN and star-forming galaxies. On the other hand, passive galaxies have W1-W2 and W2-W3 colors close to 0. 

In the paragraphs above we have analyzed the discriminating power that the mid-IR colors can have in the activity classification of a galaxy. However, the mid-IR color diagnostic tools that have been developed so far often suffer from dust obscuration effects. A well-known example is that a starburst galaxy can mimic an AGN galaxy \citep[e.g.,][]{2016ApJ...832..119H}. The introduction of an optical color is able to identify these cases and breaks the degeneracy present in the mid-IR color space. In Fig. \ref{fig:ft_dist} we see that the distribution of the SF galaxies have \textit{g}-\textit{r} color that is clearly separated from the other activity classes.

Our results (Sec. \ref{sec4}) show that the random forest diagnostic achieves an overall accuracy of 81\%. As this score was calculated on an independent sample (galaxies that were not used for its training) it is a good estimation of its general performance. In addition, the low standard deviation of the accuracy indicates that our diagnostic has stable performance.

Now, considering the above-mentioned trend, we can look back at the feature importance (Sec. \ref{sec42}) to see why some of the features were more important than others for the training of the algorithm. Observing Fig. \ref{fig:ft_imp} we see that the feature with the highest impact is the W2-W3 color. This can be explained by the strong PAH emission of star-forming galaxies which dominates the emission in the W3 (centered at 12 $\mu$m) band.

\subsection{Probability distributions} \label{sec52}

Besides the classification of each galaxy, the random forest algorithm can also give an estimation of the probability of an object belonging to each one of the classes individually. A significant difference between the probability of the first ranking and the probability of the second-ranking class indicates a highly confident classification. The probabilities we examine in this section are the calibrated predicted probabilities (see Sec. \ref{calcl}). 

\begin{sidewaysfigure*}
\centering
\includegraphics[scale=0.18]{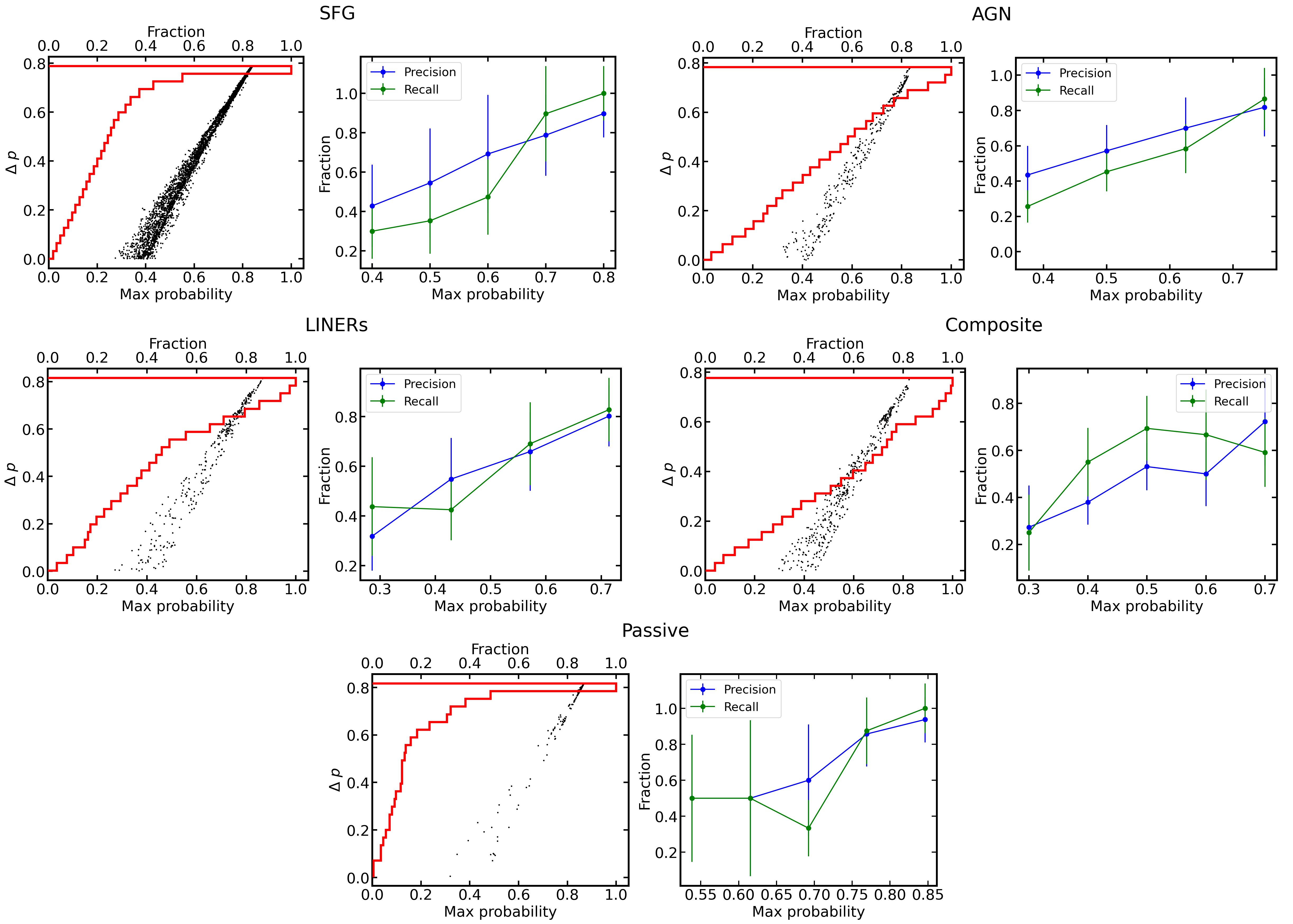}
\caption{Plots for the reliability analysis of the predicted probabilities of each galaxy activity class. For each class, we plot two diagrams. First, on the left under each class label, we plot the probability difference of the maximum and second-highest predicted probability ($\Delta p$) against the maximum predicted probability. In every such plot, each black dot represents a galaxy while the red line is the normalized cumulative histogram with respect to $\Delta p$ (top x-axis tick marks represent the fraction of the total objects). Second, on the right plot under each class label, we also plot the "recall" and "precision" scores as a function of the maximum predicted probability for each activity class.}
\label{fig:prob}
\end{sidewaysfigure*}

In order to evaluate the confidence of the classifications performed for each galaxy, we compare the probability of the highest and the second-higher ranking class. For this reason, in Fig. \ref{fig:prob}, we calculate the probability difference between the first and the second-higher probabilities, $\Delta p$, and we plot it against the maximum predicted probability for each of the five classes. Objects appearing in the top right corner of that plot have high probability to belong to the first-rank class (close to 1) while the probability difference from the second class is also high. These objects have been classified with very high confidence and thus have high reliability. Another test we perform to check the reliability of the predicted probabilities is to plot the "recall" and the "precision" for each class as a function of their predicted probability. The process to calculate these curves is the following: after we have taken objects that belong to only one class, we split them into bins based on their predicted probability. Then, we calculate the fraction of objects that have been correctly predicted to belong to the class under examination to the total predictions made in that bin to belong to that class (i.e., a measure of precision). Also, we calculate the fraction of galaxies in each probability bin that have been identified correctly to belong to a class by the new diagnostic to the total number of objects that truly belong to that class (i.e., a measure of recall). These plots are also shown in Fig. \ref{fig:prob}. The error bars that are displayed in the above-mentioned fractions ("recall" and "precision") are proportional to the square root of the instances found in each bin. So, through error propagation, we have that the error for the "recall" is $error = \sqrt{\frac{n^{2}}{N^{3}}+\frac{n}{N^{2}}}$, where n is the true positive and $N$ is the sum of the true positive and false negative examples in each predicted probability bin, while for the "precision" we use the same equation, but the n is the true positive and $N$ is the sum of the true positive and false positive examples in each predicted probability bin. By inspecting Fig. \ref{fig:prob} we see that $\Delta p$ is high for almost all classes indicating a highly confident classification. Also, we see that as the predicted probability increases the recall and precision increase, indicating the reliability of the classifications as a function of the maximum predicted probability.

By inspecting the probabilities for each class individually, we deduce that especially for the classes of star-forming and passive galaxies, the combination of high recall and high classification probabilities makes it a highly confident classifier. However, despite the excellent results for these two classes in the case of AGNs we see moderate performance scores. As shown in Sec. \ref{sec4}, regarding the predictions on the true AGN (ground truth) there was considerable confusion with the class of composite and LINER galaxies. This is because the AGN galaxies share common properties with the LINER but also with the composite galaxies. For example composite galaxies are the result of AGN activity superimposed on a star-forming component \citep{2001ApJ...556..121K} or a star-forming component with photoionization by old stellar populations \citep{2010MNRAS.403.1036C}. Similarly, old stellar populations \citep{2008MNRAS.391L..29S} or AGN could be the excitation in LINERs \citep{2009A&A...506.1107G}. This is reflected in the feature distributions we presented in Fig. \ref{fig:ft_dist}.

Another reason for this is the fact that the spectroscopic classifications that we considered as true (ground truth) are subject to aperture effects. \cite{2014MNRAS.441.2296M} studied the effect of a changing aperture on the classification of an AGN galaxy finding that AGN features change as a function of the physical distance of the region within the spectral aperture and hence as a function of the observed distance. For example, an AGN galaxy observed with an increasing aperture (starting from the core) tends to move towards the \ion{H}{II} region in a BPT diagram. Two techniques to mitigate this problem are the definition of diagnostics in a specific redshift range and the star-light subtraction but this behavior is not fully removed. Another source of aperture effects includes the difference between the optical spectra and the IR photometry (\textit{WISE}) we used for the definition of our diagnostic. The latter is discussed in the next section in detail.

In an attempt to explore the role of aperture effects for the optical colors, we explored two distinct diagnostic schemes based on the redshift. We created two separate classifiers each one specialized in a specific redshift area. One classifier was trained in the range $0.02<z<0.05$ and the other is $0.05<z<0.08$. Also, we tried a unified scheme that contained galaxies across the entire range of the redshift ($0.02<z<0.08$) with the addition of the redshift as an extra discriminating feature to the three originally considered. Both attempts failed to improve the performance. 

\subsection{Mixing between classes} \label{secm53}

In Sec. \ref{sec4}, we measured the performance of the diagnostic based on its predictions on the test subset of galaxies. We find that the classifier has excellent performance for the star-forming and passive galaxies and good performance on the rest of the classes. But, besides its high performance for star-forming galaxies, we observe that there is a non-negligible fraction, (confusion matrix; Fig. \ref{fig:cnf_mt}), of these galaxies that are predicted as composites. Composites galaxies have some common characteristics with star-forming galaxies so this is a somewhat expected behaviour. Further analysis of these misclassified SF galaxies shows that they tend to have redder \textit{g}-\textit{r} colors than a typical SF galaxy.

\begin{figure}[ht]
\begin{center}
\includegraphics[scale=0.38]{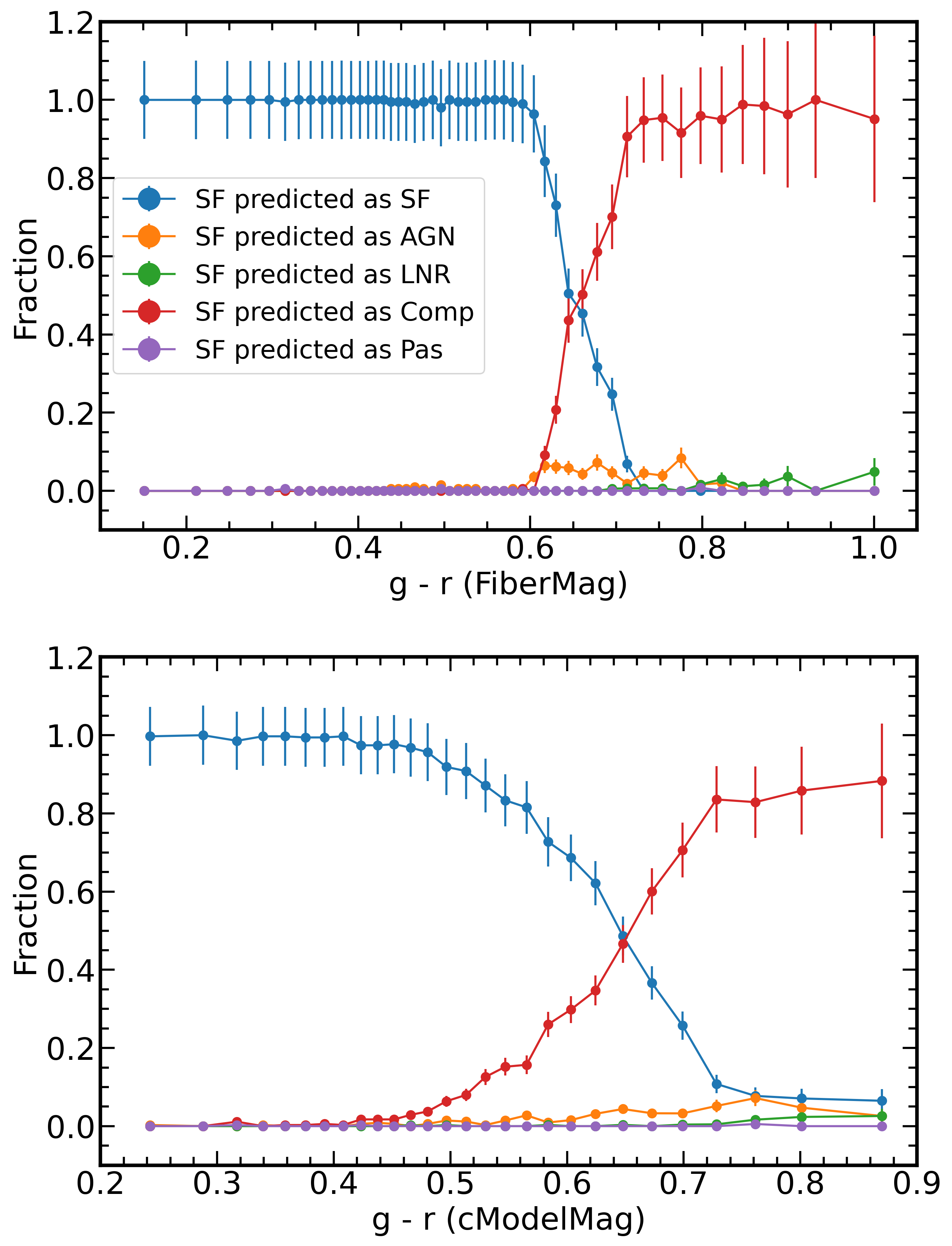}
\end{center}
\caption{Two plots of the "recall" of star-forming (SF) galaxies as a function of the \textit{g}-\textit{r} (SDSS) color. The blue line describes the fraction of SF galaxies to the total number of true SF per bin of increasing \textit{g}-\textit{r} color. The rest of the lines describe the fraction of SF galaxies that change classification (orange line, AGN; green line, LINER; red line, composite; purple, passive) to the total number of SF galaxies in a particular bin. On the top, we use the \textit{g}-\textit{r} color from the fiberMag while on the bottom plot we use the \textit{g}-\textit{r} color from the cModelMag photometry. Labels: SF, star-forming; LNR, LINER; Comp, Composite; Pas, passive.}
\label{fig:frac_gr}
\end{figure}

Figure \ref{fig:frac_gr} shows the fraction of the correctly identified star-forming galaxies (i.e., "recall") and the true SF (spectroscopic classification) that are predicted as different galaxy classes as a function fiberMag 
(\textit{g}-\textit{r}) color (top panel) which is used for the classification (Sec. \ref{sec34}) and the cModelMag (\textit{g}-\textit{r}) color which reflects the overall light of a galaxy. We estimate the errors as described in Sec. \ref{sec52}. It is clear that for SF galaxies with bluer \textit{g}-\textit{r} colors, the classifier has a "recall" rate close to 1. On the other hand, as the optical colors of the SF galaxies become redder, their recall drops, and the galaxies are predicted almost exclusively as composites. This result indicates that as these galaxies have gradually older stellar populations their IR colors resemble these of a composite galaxy. Another interesting fact is that the "recall" of the star-forming galaxies is more gradual if we use the integrated photometry to calculate the \textit{g}-\textit{r} color, suggesting that star-forming galaxies with a prominent bulge are more likely to be classified as composite. However, the value of the \textit{g}-\textit{r} color at which the fraction of the SF galaxies predicted as SF becomes equal to the fraction of SF galaxies predicted as composites remains almost the same for both photometries (\textit{g}-\textit{r} $\sim 0.65$). This suggests that the classification of a star-forming galaxy is insensitive to aperture effects as star formation is a galaxy-wide phenomenon and is not concentrated in one area as the \ion{H}{II} regions are scattered across the galaxy disk.

To further analyze the performance of the classifier we have to understand the underlying activity of each class in more detail. Starting from AGN galaxies, their emission comes from accretion of circumnuclear material onto the SMBH at their cores. The energy source of a LINER galaxy can be either a SMBH or old stellar populations, including post-AGB stars \citep{1994A&A...292...13B,2008MNRAS.391L..29S,2013A&A...555L...1P}. Composite galaxies may have some star-formation activity but can also harbor an accreting SMBH. Lately, it has been suggested that these galaxies may host old stellar populations that can actively contribute to their emission-line spectrum \citep{2008MNRAS.391L..29S}. Finally, a passive galaxy is a system without any star formation or AGN activity, with low to no reservoirs of dust and cold gas, with its main component being the old stellar populations.

The class of passive galaxies is very well characterized by this diagnostic tool. However, the confusion matrix (Fig. \ref{fig:cnf_mt}) shows that some passive galaxies are misclassified by the diagnostic as LINERs. This is consistent with studies claiming that LINER-like activity originates from old stellar populations like a passive galaxy \citep{2019AJ....158....2B}. For the class of LINERs, we see that most of the misclassified galaxies are predicted as passive (in agreement with the connection between the LINER-like spectra and old stellar populations) followed by the composite, and finally the AGN. Composite galaxies often have large bulges \citep{2000A&A...355..949F,2001A&A...376..878O} and they contain old stellar populations similar to passive galaxies. Therefore, composite galaxies could be excited by old stellar populations as discussed earlier, and their photometric colors can resemble those of passive galaxies (Fig. \ref{fig:ft_dist}). Finally, LINER activity can sometimes be attributed to an active nucleus \citep[i.e.,][]{1999AdSpR..23..813H}. In this case, the optical and IR colors of the galaxy would be consistent with those of AGNs. For the class of composite galaxies, we see that (Fig. \ref{fig:cnf_mt}) the misclassified galaxies are mainly predicted as AGNs, which is an expected result considering all the above-mentioned facts in this section.

\begin{figure}[h]
\begin{center}
\includegraphics[scale=0.47]{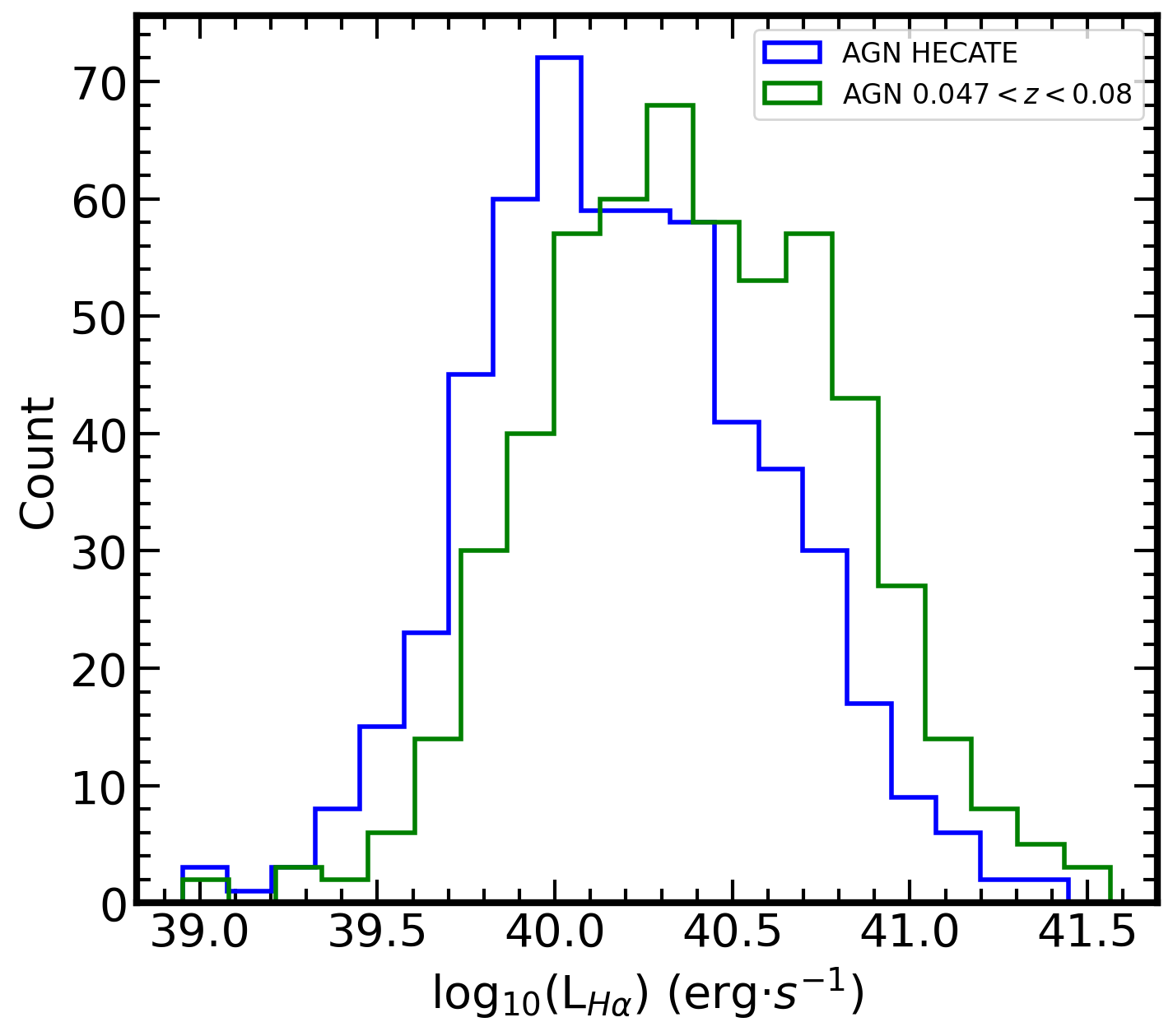}
\end{center}
\caption{Histogram of the H$\alpha$ luminosity for our sample of spectroscopically classified AGN galaxies (considered as ground truth). We split them into two redshift bins. The first bin is from $z=0.02$ to $z=0.05$ \citep[HECATE catalog;][]{2021MNRAS.506.1896K}, plotted with the blue line and in the second is from $z=0.05$ to $z=0.08$, plotted with the green line.}
\label{fig:ha_lum}
\end{figure}

\begin{figure}[h]
\begin{center}
\includegraphics[scale=0.4]{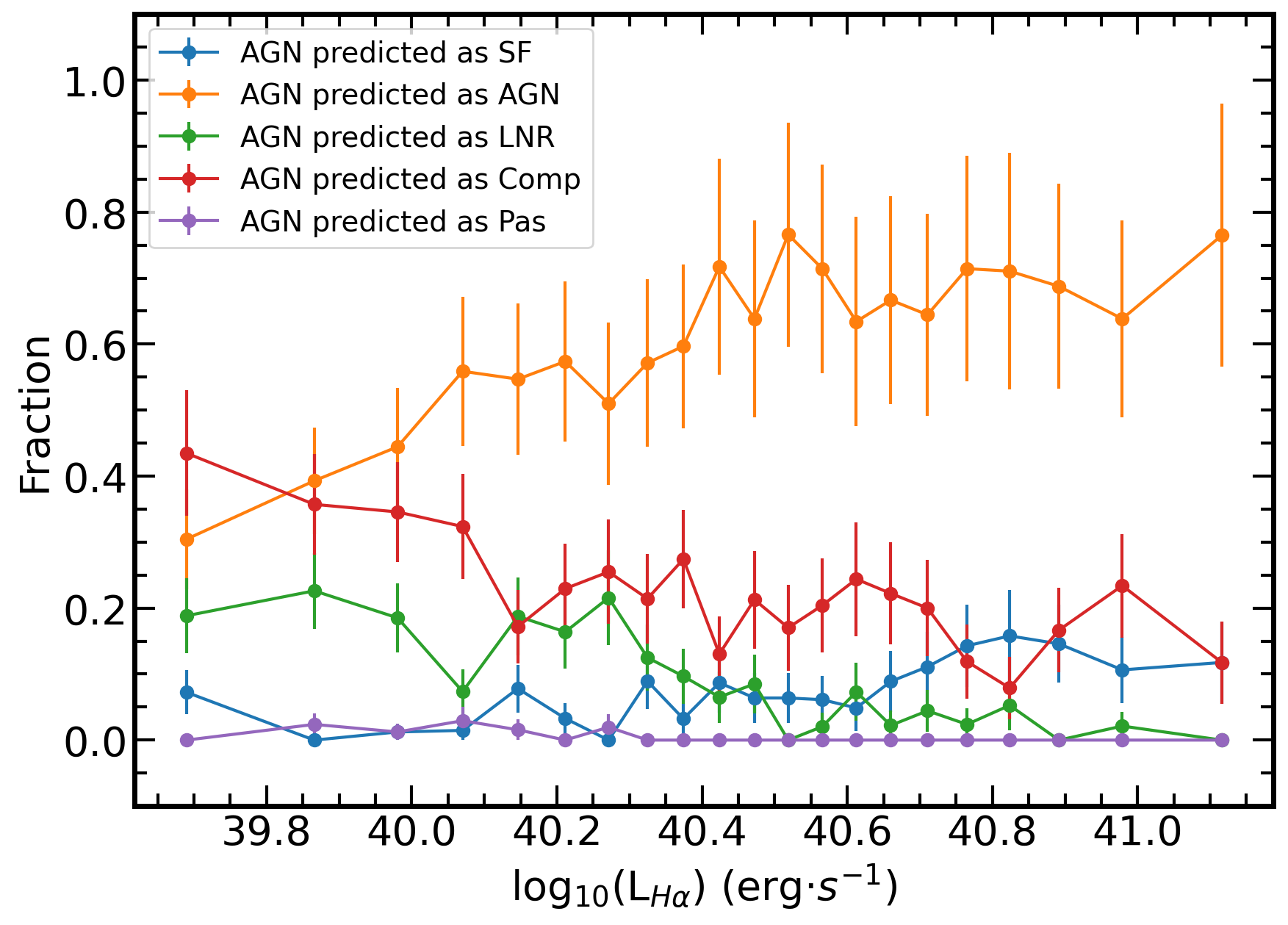}
\end{center}
\caption{The fraction of the correctly identified AGN (true positives) to the total true AGN (i.e., a measure of "recall" or "completeness') as a function of the AGN H$\alpha$ luminosity for all spectroscopically selected AGN galaxies in our sample (orange line). All other colored lines represent the fraction of true AGN galaxies that the diagnostic predicted to belong to a class other than the AGN (false negatives) to the total true AGN (blue, SF; green, LINER; red, composite; purple, passive). All fractions are calculated after the galaxies have been split into bins of increasing H$\alpha$ luminosity containing the same number of galaxies. Labels: same as Fig. \ref{fig:frac_gr}.}
\label{fig:frac_ha_lum}
\end{figure}

During the evaluation of our model we found that the "recall" of the AGN galaxies tends to increase with increasing distance (Fig. \ref{fig:rcl_cl_z}). An explanation of this effect is that due to the aperture effects (as well as volume and sensitivity effects) the AGN identified at larger distances tend to be more luminous. This is seen in Fig. \ref{fig:ha_lum} which shows the histograms of H$\alpha$ luminosity of the AGNs galaxies for two redshift bins, one for the very nearby galaxies ($0.02<z<0.05$) and one for galaxies further away ($0.05<z<0.08$). More luminous AGNs are more likely to dominate the mid-IR colors of a galaxy. Further supporting evidence is provided by Fig. \ref{fig:frac_ha_lum} which shows the fraction of spectroscopically selected AGNs classified in different classes by the classifier as a function of their H$\alpha$ luminosity. To produce this plot we split the AGNs (regardless of redshift) in our sample into bins of increasing H$\alpha$ luminosity. Then, we apply our diagnostic and calculate the fraction of objects predicted to belong to each class with respect to the spectroscopic AGN in each bin. We can see that the fraction of the correctly identified (i.e., the "recall") AGN increases as their H$\alpha$ luminosity increases. It is also clear that the diagnostic confuses cases of low-luminosity AGNs as composites, which is reasonable if we also consider the aperture effects. Based on this we estimate that AGNs with H$\alpha$ luminosities below $\sim 5 \times 10^{40}$ erg$\cdot$s$^{-1}$ are increasingly missed to composite galaxies. 

\begin{figure*}
\begin{center}
\includegraphics[scale=0.35]{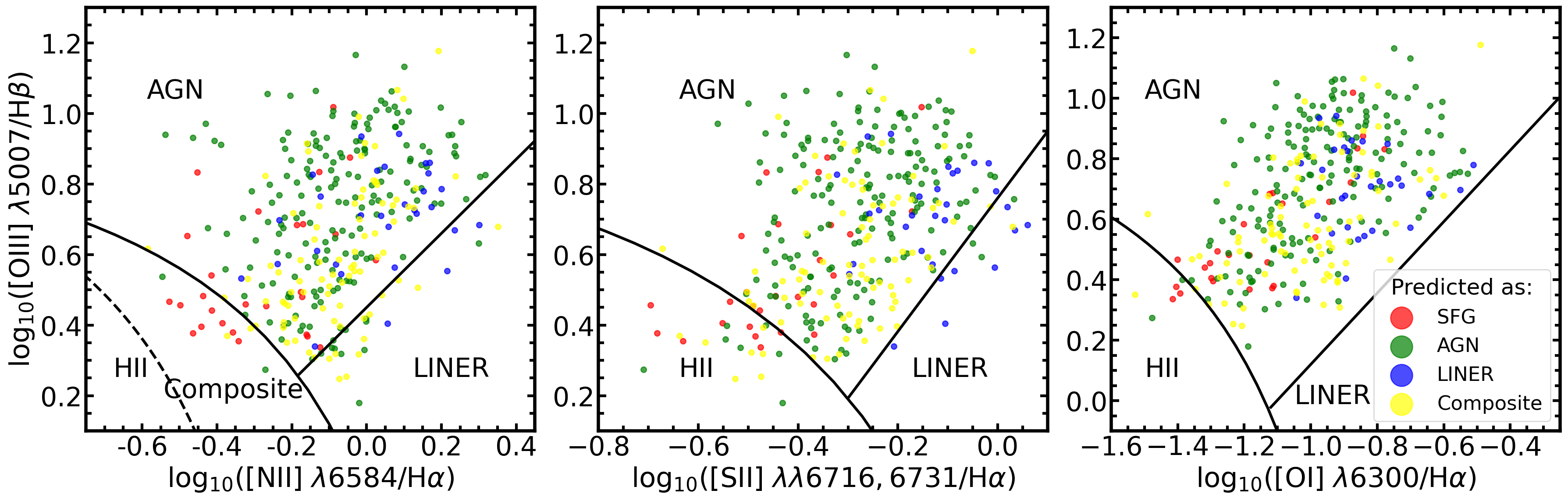}
\end{center}
\caption{Diagrams of [\ion{O}{III}] $\lambda$5007/H$\beta$ against [\ion{N}{II}] $\lambda$6584/H$\alpha$ (left), [\ion{S}{II}] $\lambda\lambda$6716,6731/H$\alpha$ (middle), and [\ion{O}{I}] $\lambda$6300/H$\alpha$ (right) showing the location of an optically selected sample of AGNs from our test sample (Sec. \ref{sec41}). The points are color-coded depending on their classification based on our diagnostic: AGN are the green dots, SF are the red dots, LINERs are blue dots, and composites are the yellow dots. We see that, since these are two-dimensional projections of the four-dimensional space used for the optical line-ratio classification, some AGN may fall outside the AGN demarcation line. The black solid curve is the extreme starburst line defined by \cite{2001ApJ...556..121K}. The straight black line is the separating line between AGN and LINER galaxies as defined by \cite{2007MNRAS.382.1415S}. The black dashed curve is the \cite{2003MNRAS.346.1055K} line separating star-forming from composite galaxies.}
\label{fig:BPT_AGN_pred}
\end{figure*}

Another interesting fact about the class of AGN galaxies comes from the misclassification instances that our diagnostic tool makes. In Fig. \ref{fig:BPT_AGN_pred} we plot the emission line ratios of [\ion{O}{III}]/H$\beta$ against [\ion{N}{II}]/H$\alpha$, [\ion{S}{II}]/H$\alpha$, and [\ion{O}{I}]/H$\alpha$. The location of spectroscopic AGN classified in different classes on the [\ion{O}{III}]/H$\beta$ against [\ion{N}{II}]/H$\alpha$ diagram shows that the AGN galaxies that have been misclassified as SF are located primarily close to the line of maximum starburst defined by \cite{2001ApJ...556..121K} which is the line that separates AGN and SF galaxies. In addition, in the plot of [\ion{O}{III}]/H$\beta$ against [\ion{S}{II}]/H$\alpha$, we see that AGN galaxies that have been predicted as LINERs are located very close to the separating line of \cite{2007MNRAS.382.1415S} that separates AGN and LINER galaxies and are systematically located in the upper right area of the AGN locus having higher values of [\ion{S}{II}]/H$\alpha$. Also, we see that misclassified AGNs predicted as LINERs have similar trend in the plot of [\ion{O}{III}]/H$\beta$ against [\ion{O}{I}]/H$\alpha$ as in the [\ion{O}{III}]/H$\beta$ against [\ion{S}{II}]/H$\alpha$. However, we notice that there is significant mixing between the composite and AGN galaxies. The misclassified AGN predicted as composites have no specific trend as they are scattered across the AGN locus of these plots, an effect that can be attributed to the use of the galaxy-wide IR colors. Nonetheless this is acceptable, especially considering the flexibility provided by using the integrated colors of the galaxies and the excellent performance in the case of the other classes.

\subsection{Comparison with other methods} \label{comp_mat}

\begin{figure*}[ht]
\begin{center}
\includegraphics[scale=0.43]{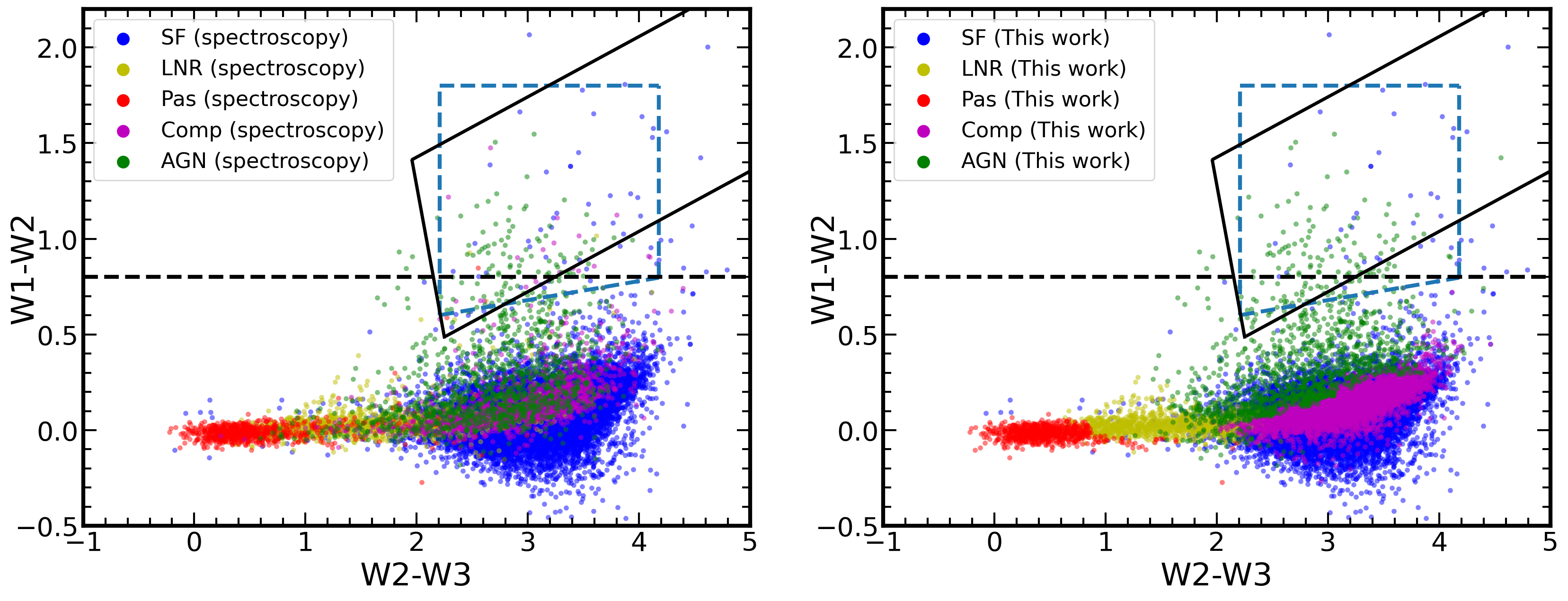}
\end{center}
\caption{Color-color plots of the W1-W2 against the W2-W3 for our sample of galaxies. On the left, we plot the galaxies based on their spectroscopic classification (true class). On the right, we see the same plot for the same sample of galaxies but the class labels of each galaxy have been assigned by the new diagnostic tool. The solid black line is the locus of AGN galaxies as it was defined by \cite{2012MNRAS.426.3271M}, while the black dashed line is the demarcation line between an AGN and a non-AGN galaxy defined by the \cite{2012ApJ...753...30S}. The blue dashed lines define the AGN selection box as defined by \cite{2011ApJ...735..112J}. SF galaxies are represented with blue, LINERs with yellow, passive with red, composite with purple, and AGN with green points. Labels in the legend are the same as in Fig. \ref{fig:frac_gr}. We see that there is a significant population of spectroscopic AGN that is located below the existing IR diagnostics which are correctly identified with our diagnostic. There is also a population of extreme star-forming galaxies that is located in the AGN locus of the existing AGN diagnostics which is also correctly classified by our diagnostic.}
\label{fig:mt_RF}
\end{figure*}

In order to determine if this new diagnostic, which is based on infrared and optical colors, provides any advantage over the already established ones we compare their performances against the performance of our diagnostic. Taking into account that they have been based on different criteria and parent samples we discuss their advantages and disadvantages.

A widely used infrared diagnostic is the W1-W2 $\geq$ 0.8 criterion \citep{2012ApJ...753...30S,2013ApJ...772...26A}, where a demarcation line based on two \textit{WISE} bands (band 1 and 2) separates AGN galaxies from the rest of the galaxies. Other similar diagnostic tools have been introduced by \cite{2011ApJ...735..112J},\cite{2012MNRAS.426.3271M}, and \cite{2014RMxAA..50..255C}. The first two are two-dimensional diagnostics defined based on the W1-W2 against the W2-W3, while the latter is based on a plot of W3-W4 against W2-W3 colors. In the case of the \cite{2012MNRAS.426.3271M} diagnostic tool, which focuses on high-luminosity AGNs, the authors define an AGN selection wedge on the upper right corner of the plot defined by the equations: $(W1-W2) = 0.315*(W2-W3) + 0.796$, $(W1-W2) = 0.315*(W2-W3) - 0.222$, and $(W1-W2) = -3.172*(W2-W3) + 7.624$. To test the applicability of this diagnostic to the wider population of (non-X-ray selected) AGNs, in Fig. \ref{fig:mt_RF} we plot the W1-W2 against the W2-W3 colors of the galaxies in our full sample (Sec. \ref{sec24}) color-coded according to their spectroscopic classification (left) and the classification based on our diagnostic (right). The galaxies presented in Fig. \ref{fig:mt_RF} originate from the SDSS sample in the redshift range of $z=0.02$ to $z=0.08$. We see that a significant number of AGN are located outside of the AGN locus of \cite{2012MNRAS.426.3271M} and below the demarcation line of \cite{2012ApJ...753...30S}. This behavior holds even with the diagnostic of \cite{2011ApJ...735..112J}, since all three diagnostics \citep{2011ApJ...735..112J,2012MNRAS.426.3271M,2013ApJ...772...26A} are based on luminous AGN samples for their definition. Furthermore, we observe that there are star-forming galaxies that the two methods of AGN identification \citep{2012MNRAS.426.3271M,2013ApJ...772...26A} classify wrongly as AGN which we discuss further in Sec. \ref{sec55}. 

\begin{table}[ht]
\centering
\caption{Comparison of the classification results between our diagnostic and other widely used mid-IR diagnostics.}
\begin{tabular}{|c|c|c|c|} 
\hline\multicolumn{2}{|c|}{\diagbox{Other \\ diagnostics~}{This work}} & ~AGN~ & Non-AGN  \\ 
\hline
\hline
\multirow{2}{*}{JR11} & AGN & 25 (7.6\%) & 1 (0.3\%)\\ 
\cline{2-4}
 & Non-AGN & 158 (48.0\%) & 145 (44.1\%)\\
\hline
\hline
\multirow{2}{*}{MT12} & AGN  & 22 (6.7\%)& 1 (0.3\%)\\ 
\cline{2-4}
 & Non-AGN & 161 (48.9\%) & 145 (44.1\%)\\ 
\hline
\hline
\multirow{2}{*}{CZ14} & AGN & 85 (31.4\%) & 49 (18.0\%) \\ 
\cline{2-4}
 & Non-AGN & 73 (26.9\%) & 64 (23.6\%) \\
\hline
\hline
\multirow{2}{*}{\begin{tabular}[c]{@{}c@{}} AS18 \\(R75)\end{tabular}} & AGN & 55 (16.7\%) & 2 (0.6\%)\\ 
\cline{2-4}
 & Non-AGN & 128 (38.9\%)& 144 (43.8\%)\\ 
\hline
\hline
\multirow{2}{*}{\begin{tabular}[c]{@{}c@{}} AS18 \\(R90)\end{tabular}}  & AGN & 32 (9.7\%) & 1 (0.3\%) \\ 
\cline{2-4}
 & Non-AGN & 151 (45.9\%) & 145 (44.1\%) \\ 
\hline
\end{tabular}
\label{tab:agn_comp}
\tablefoot{The comparison is performed on an optically selected sample of 329 AGN galaxies. Columns represent our classification results while rows are the classification obtained by the other diagnostics. The R75 and R90 refer to the two available schemes in the work of \cite{2018ApJS..234...23A} that give 75\% and 90\% reliability for AGN selection respectively. References: JR11, \cite{2011ApJ...735..112J}; MT12, \cite{2012MNRAS.426.3271M}; CZ14, \cite{2014RMxAA..50..255C}; AS18, \cite{2018ApJS..234...23A}.}
\end{table}

Focused on the AGN galaxies we include one more mid-IR diagnostic and we perform a quantitative comparison between our diagnostic tool and the rest widely used infrared diagnostic tools that were mentioned earlier. The other diagnostic we include in our comparison was defined by \cite{2014RMxAA..50..255C}. This particular diagnostic focuses on spectroscopically selected low-luminosity AGNs (LLAGNs). The selection criteria of this diagnostic consist of a 2-dimensional plot of W3-W4 against W2-W3 that is separated into four parts with two crossing lines; $(W3-W4) = 1.6*(W2-W3) + 3.2$ and $(W3-W4) = -2.0*(W2-W3) + 8.0$.

To obtain a quantitative comparison between our diagnostic and the four aforementioned tools we use the test sample (Sec. \ref{sec41}) and choose only all the AGN galaxies (329 in total), which will be considered as the ground truth for the comparison. We focus on AGN since these tools are tailored for the classification of AGN in which our diagnostic shows weaker performance. Then we apply our new diagnostic and the diagnostic tools of \cite{2012MNRAS.426.3271M}, \cite{2018ApJS..234...23A}, \cite{ 2014RMxAA..50..255C}, and \cite{2011ApJ...735..112J} on that sample of optically selected AGN galaxies. Since the other diagnostics offer only AGN or non-AGN classification we adapt our results accordingly. Any galaxy that receives a classification by our diagnostic other than AGN (SF, LINER, composite, or passive) is characterized as non-AGN. In Table \ref{tab:agn_comp} we present the results of this comparison between the classifications made by our diagnostic and the other diagnostic methods. We see that despite our moderate scores for the class of AGN galaxies our diagnostic tool identifies more AGN galaxies than any other method tested here.

\begin{figure}[h]
\begin{center}
\includegraphics[scale=0.48]{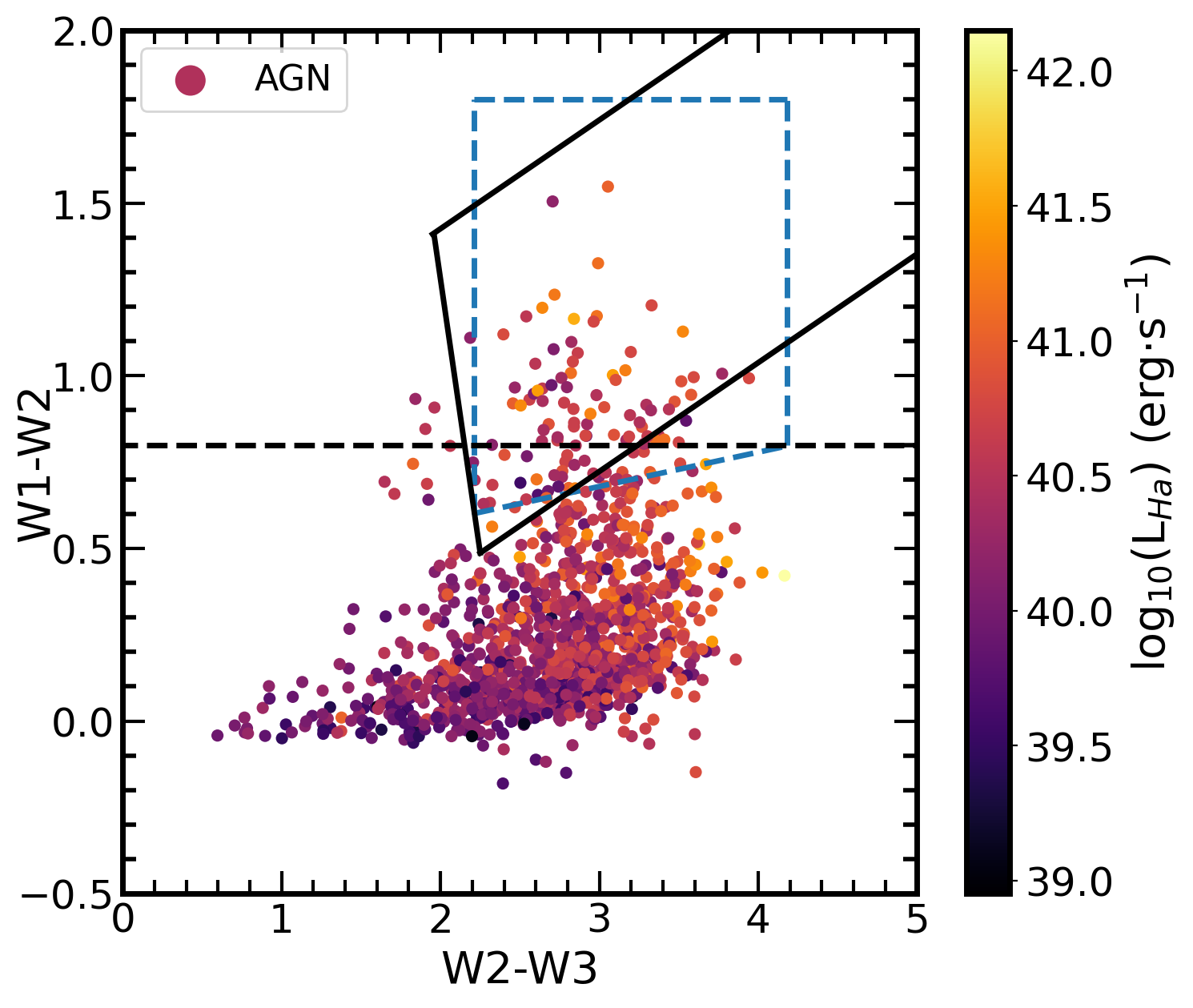}
\end{center}
\caption{The W1-W2 color against the W2-W3 color plot for the spectroscopically classified AGNs in our sample color-coded with H$\alpha$ luminosity. The solid black line is the locus of AGN galaxies as it was defined by \cite{2012MNRAS.426.3271M}, while the black dashed line is the demarcation line between an AGN and a non-AGN galaxy defined by the \cite{2012ApJ...753...30S}. The blue dashed lines define the AGN selection box as defined by \cite{2011ApJ...735..112J}.}
\label{fig:agn_lum}
\end{figure}

While the existing diagnostics are very effective in identifying reliable samples of AGNs (albeit with some contamination by extreme starburst; see next section), they are biased towards the more obscured and more luminous AGNs, missing the bulk of their populations. In Fig. \ref{fig:agn_lum} we plot the W1-W2 against the W2-W3 for all the spectroscopically classified AGNs in our sample color-coded with their H$\alpha$ luminosity. In that figure we see that only the more luminous AGNs will be selected by the diagnostics of \cite{2011ApJ...735..112J,2012MNRAS.426.3271M, 2012ApJ...753...30S}. Instead, our diagnostic also provides samples of lower-luminosity AGN, with the unavoidable mixing with composite and LINER galaxies. Furthermore, since they are driven by the classification of one activity class they are not efficient in identifying reliable samples of star-forming galaxies (which have strong contamination by lower-luminosity AGN) or other types of galaxies (composite, passive, and LINER).

\subsection{Star-forming galaxies with extremely red mid-IR colors} \label{sec55}

By observing more closely Fig. \ref{fig:mt_RF} (left panel), we see that there is a significant number of spectroscopically classified star-forming galaxies with mid-IR colors of W1-W2 $\geq$ 0.8 \citep{2012ApJ...753...30S}. Normally these galaxies would have been classified as AGN by most mid-IR selection methods \citep{2011ApJ...735..112J,2012MNRAS.426.3271M,2013ApJ...772...26A}. In contrast to these AGN selection methods, our diagnostic tool is able to separate these cases as we can see that it manages to retrieve $\sim$82\% of the spectroscopically classified star-forming galaxies that are located above the \cite{2012ApJ...753...30S} line. This means that our diagnostic has the ability to correctly identify a case where a starburst galaxy looks like an AGN \citep[e.g.,][]{2016ApJ...832..119H}. This ability is the result of the use of optical colors which is a tell-tale signature of extreme starburst galaxies.

By further investigating the optical spectra and SDSS images of these peculiar star-forming galaxies (W1-W2 $\geq$ 0.8) we find that their majority have spectra that are indicative of an \ion{H}{II} region appearing as blue compact spherical objects in the SDSS images. There is also a population with redder SDSS colors, indicating the presence of dust. These star-forming galaxies appear to have W2-W3 colors that are systematically redder than the AGNs. An interesting fact is that the "dusty" star-forming galaxies are mainly located in the area of W2-W3 $\lesssim$ 3.5, while above that value almost all SF galaxies seem to be blue and compact.

Additional evidence for a population of SF galaxies contaminating the mid-IR AGN diagnostics is provided by the \cite{2016ApJ...832..119H}. The focus of this work is the properties of a spectroscopically selected sample of dwarf galaxies including AGN and star-forming galaxies. They find that a significant fraction of the optically selected AGN galaxies is not selected as AGNs by the mid-IR diagnostics of \cite{2011ApJ...735..112J} and \cite{2012ApJ...753...30S}. They also find that in these two AGN diagnostics, there is a significant contamination in the mid-IR selected samples of AGNs, a result that is supported by our analysis.

These galaxies with AGN-like mid-IR colors are consistent with the ''blueberry'' galaxies. The latter are characterized by compact sizes, extreme blue colors, and \ion{H}{II} region-like spectra \citep{2017ApJ...847...38Y}. Also, these galaxies tend to be metal-poor with extreme mid-IR colors \citep{2016ApJ...832..119H}. In our analysis, we find that galaxies with extreme mid-IR colors (W2-W3 $\gtrsim$ 3.5) that are spectroscopically identified as star-forming have \textit{g}-\textit{r} color $< 0$ (median $\sim -0.35$). In fact, this is the main differentiating factor between the AGNs with extreme mid-IR colors and the blueberry galaxies for our diagnostic. Star-forming galaxies above the W1-W2 = 0.8 and with W2-W3 $\lesssim$ 3.5 seem to have optical \textit{g}-\textit{r} colors redder than the ones with W2-W3 $\gtrsim$ 3.5 (median $\sim 0.2$).

\cite{2017ApJ...847...38Y} provide a catalog of 41 objects with a spectroscopically selected sample of blueberry galaxies that appear to be compact and blue in the SDSS $gri$ (Blue-Green-Red) images. After cross-matching these galaxies with the HECATE catalog \citep{2021MNRAS.506.1896K}, value-added catalog for the local Universe (distances up to 200 Mpc) and applying quality cuts to photometry (as described in Sec. \ref{sec24}), we apply our diagnostic. We found that 7 out of 41 objects satisfied our quality criteria. All 7 of them have been classified correctly by the diagnostic as being star-forming, while 4 out of 7 of them are above the AGN demarcation line of \cite{2012ApJ...753...30S}. Also, we find that the \cite{2011ApJ...735..112J} wedge classifies as AGN 2 out of 7 objects while \cite{2012MNRAS.426.3271M} classifies 3 out of 7 as being AGN. Of course, these results must be taken with caution as the number of objects is limited.

\section{Conclusions} \label{concl}

In this work, we combined mid-IR with optical photometry to define a new galactic activity diagnostic to provide a tool that includes all activity classes under one unified scheme while offering improved performance. Our results are summarised in the following points:
\begin{enumerate}
  \item In this work, we defined a new machine learning activity diagnostic tool for galaxy classification that can discriminate between five different classes of galaxies. This is the first machine-learning based tool for galaxy classification that includes not only active but also passive galaxies. The code with application instructions are available through the GitHub repository \footnote[2]{\url{https://github.com/BabisDaoutis/GalActivityClassifier}}.
  \item This diagnostic extents the existing IR diagnostics to additional types of activity, it is more sensitive to LLAGN, and it can be used for local galaxies ($z\sim0$) up to $z\sim0.08$.
  \item The addition of the optical color, \textit{g}-\textit{r}, to the two infrared \textit{WISE} colors that were used in previous works, establishes a diagnostic that can go beyond the bimodal classification (AGN or non-AGN) of most diagnostic methods allowing their extension to 4 different activity classes as well as, passive galaxies.
  \item There is some mixing between the classification of some objects, but this is observed mainly in classes that share common properties (e.g., composite and AGN galaxies). Although this results in reduced performance scores for the class of AGN galaxies, the performance we achieve is superior to that achieved by previous works. However, the addition of extra features (e.g., UV colors) may help to improve the performance, even though it will limit its applicability to objects with photometry from the UV to infrared wavelengths.
  \item By including optical information (the \textit{g}-\textit{r} color), our diagnostic is able to distinguish between a starburst galaxy with extreme mid-IR colors (e.g., blueberries) from an obscured AGN, that would have been classified as true AGNs based on traditional IR diagnostic methods.
 \end{enumerate}
There are several directions that can be taken to improve the performance of the diagnostic, especially for improving the performance of the AGNs, as well as reducing the mixing between the AGN and composite galaxies. These include: adding features that are characteristic of AGN activity (e.g., UV colors), luminosity, or even spectral information, but at the cost of reducing the applicability of the diagnostic to a larger sample of data.

\begin{acknowledgements} We thank the anonymous referee for their constructive
comments and suggestions that helped to improve this work and the clarity of this manuscript. We thank Paolo Bonfini for very useful disscusions on machine-learning methods that helped to improve the performance of the classifier. We also thank Sotiria Fotopoulou for discussions on the classification of the AGN. CD and EK acknowledge support from the Public Investments Program through a Matching Funds grant to
the IA-FORTH. The research leading to these
results has received funding from the European Research Council under the
European Union’s Seventh Framework Programme (FP/2007-2013) / ERC Grant
Agreement n. 617001, and the European Union’s Horizon 2020 research and
innovation programme under the Marie Skłodowska-Curie RISE action, Grant
Agreement n. 873089 (ASTROSTAT-II). KK is supported by the project ''Support of the international collaboration in astronomy (Asu mobility)'' with the number: CZ 02.2.69/0.0/0.0/18\_053/0016972.
Funding for SDSS-III has been provided by the Alfred P. Sloan
Foundation, the Participating Institutions, the National Science
Foundation, and the U.S. Department of Energy Office of Science.
The SDSS-III web site is http://www.sdss3.org/.
SDSS-III is managed by the Astrophysical Research Consortium
for the Participating Institutions of the SDSS-III Collaboration
including the University of Arizona, the Brazilian Participation
Group, Brookhaven National Laboratory, Carnegie Mellon University, University of Florida, the French Participation Group,
the German Participation Group, Harvard University, the Instituto
de Astrofisica de Canarias, the Michigan State/Notre Dame/JINA
Participation Group, Johns Hopkins University, Lawrence Berkeley
National Laboratory, Max Planck Institute for Astrophysics, Max
Planck Institute for Extraterrestrial Physics, New Mexico State University, New York University, Ohio State University, Pennsylvania
State University, University of Portsmouth, Princeton University,
the Spanish Participation Group, University of Tokyo, University
of Utah, Vanderbilt University, University of Virginia, University
of Washington, and Yale University.  
\end{acknowledgements}

\bibliographystyle{aa}
\bibliography{references}

\begin{appendix}

\section{Optimization of significant hyperparameters} \label{app1}

The process of optimizing this activity diagnostic is performed with the use of the \texttt{GridSearchCV} algorithm which is provided by the \texttt{scikit-learn} Python 3 package, version 1.1.2. The determination of the optimal values of each hyperparameter is usually done by training the algorithm several times with different choices of the hyperparameter values each time, typically by means of a grid search. The performance of the algorithm is evaluated at each point of the grid, and the optimal set of parameters that maximizes the performance is chosen. Since only some of the available hyperparameters have significant impact on the performance of our diagnostic, it is inefficient to perform a grid search including all of them. We find that only the following seven hyperparameters are important:
\begin{itemize}
    \item[$-$] \texttt{n\_estimators} : The number of decision trees.
    \item[$-$] \texttt{max\_depth} : The maximum depth of a decision tree.
    \item[$-$] \texttt{min\_samples\_split} : The minimum required number of samples to split a node.
    \item[$-$] \texttt{min\_samples\_leaf} : The minimum number of samples in a leaf node.
    \item[$-$] \texttt{max\_leaf\_nodes} : The maximum allowable number of terminal nodes in a tree.
    \item[$-$] \texttt{max\_samples} : The number of samples from the training set to build each tree.
    \item[$-$] \texttt{criterion} : The function to measure the quality of a split.
\end{itemize}

\begin{figure}[h]
\includegraphics[scale=0.25]{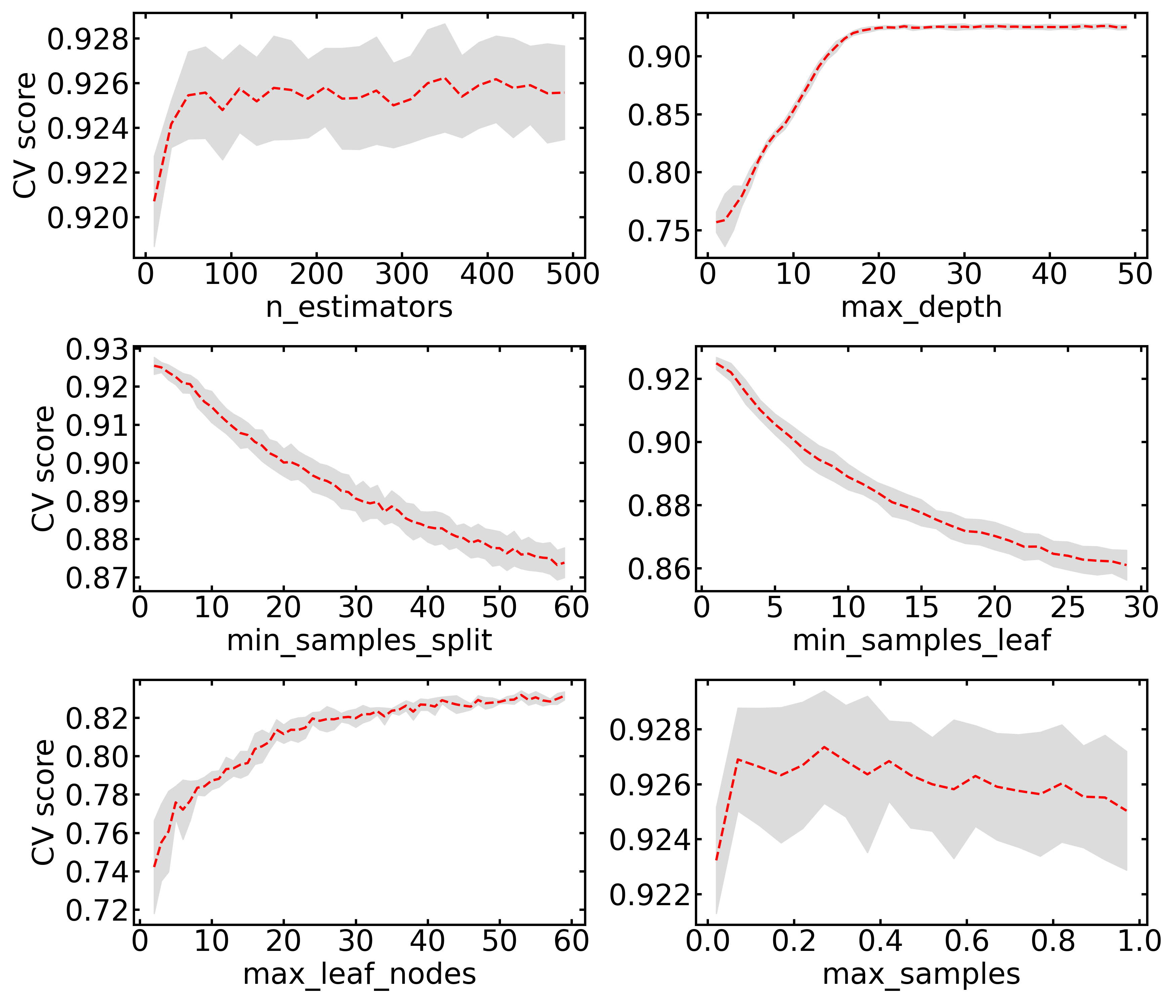}
\caption{Validation curves for the determination of the best hyperparameter ranges attempting to reduce the possible values of each hyperparameter. The CV scores refer to accuracy as a function of each hyperparameter. The red line is the average accuracy calculated with the k-fold cross-validation method (k = 5) while the shaded grey area is the standard deviation (1$\sigma$) of the k-scores for each value of the hyperparameter under examination.}
\label{fig:val_crvs}
\end{figure}

However, since a broad grid search in a 7-dimensional space can still be computationally intensive we first narrow the range of these parameters by calculating its performance for different values of each hyperparameter separately. To do this, in Fig. \ref{fig:val_crvs} we calculate the validation curves which show different performance metrics as a function of each hyperparameter value. More specifically, each validation curve has the performance score of a metric (e.g., accuracy) on the y-axis and a range of the possible values of one hyperparameter (e.g., n\_estimators) on the x-axis while keeping all the other hyperparameters constant. The performance reported on each plot is the performance on the testing fold calculated with the cross-validation (CV) method which is performed by splitting the data into k folds and using k-1 folds for its training and one for testing its performance. The algorithm is trained k times in total for each combination of hyperparameters by cycling the folds so that all k folds have been in the position of the testing fold once. The reported performance score is the average of these k scores.

By inspecting Fig. \ref{fig:val_crvs} we see that the ranges of values for the best hyperparameters are found in the areas where the scores start to converge to the best achievable score (curves start to converge parallel to the x-axis). When this happens it means that overfitting starts to occur. This way we significantly reduce the ranges of the hyperparameters that we have to test to find the optimal ones. Afterwards, from these validation curves we determine the sensitivity of the algorithm on the different hyperparameters, we find the ranges of the parameters that significantly affect its performance in order to perform a grid search around these ranges. The range for each hyperparameter was found by inspecting the behavior of the accuracy score.
\begin{table}[b]
\caption{Important hyperparameters that were optimized for the adaption random forest algorithm in our sample of galaxies.}
\centering
\resizebox{\columnwidth}{!}{
\begin{tabular}{lcc}
\hline\hline
Parameter & Search range & Best value  \\
\hline
\texttt{n\_estimators} & 10-250 & 120 \\
\texttt{max\_depth} & 15-20 & 17  \\
\texttt{min\_samples\_split} & - & 'default' \\
\texttt{min\_samples\_leaf} & - & 'default' \\
\texttt{max\_leaf\_nodes} & 25-70 & 30  \\
\texttt{max\_samples} & 0.1-0.9 & 0.6 \\
\texttt{class\_weight} & - & 'balanced\_subsample' \\
\texttt{criterion} & - & 'Gini' \\
\hline
\end{tabular}
}
\label{tablehyp}
\end{table}

The next step is to use these best value ranges for each hyperparameter extracted by the plots as input to the grid search algorithm. Then, that algorithm will make combinations of the hyperparameters from the best value ranges and it will fit each derived model in order to find which model has the best performance scores. In Table \ref{tablehyp} we present the hyperparameters that have been considered for optimization along with their search ranges and their best values. 

After we have obtained the set of the best hyperparameters we can test the stability of the algorithm when it is trained and tested on all possible subsets for the training and the test data. One such performance stability test is to perform k-fold cross-validation and observe the change in the accuracy score. A stable algorithm should have a low standard deviation on its accuracy which means that the performance of the algorithm does not significantly fluctuate between its subsequent application on similar data, although small fluctuations are unavoidable as a result of the stochastic nature of the algorithm. Even though the sample is fairly large, the data for some minority classes are low. For that reason, the number of folds chosen here is 5 (each fold consists of 8190 objects), as even though we have a significant number of objects, some classes will be underrepresented with a choice of a higher number of folds. The overall accuracy score of the model when calculated with the k-fold cross-validation method is 81\%$\pm$1\%. This suggests that the fluctuations of the performance scores are low, suggesting a stable algorithm.

\end{appendix}
\end{document}